\documentclass[aps,prd,twocolumn,superscriptaddress,floatfix,showpacs,amsmath,letterpaper]{revtex4-1}

\usepackage{lineno}
\usepackage{graphicx} 
\usepackage{dcolumn} 
\usepackage{bm} 
\usepackage{epstopdf} 
\bibpunct{[}{]}{,}{n}{}{} 
\usepackage{hyperref}

\newcommand {\MM} [1] {\ensuremath{#1}}
\newcommand {\tsp} [1] {\ensuremath{\mskip #1\thinmuskip}}
\newcommand {\IT} [1] {\ensuremath{#1}}
\newcommand {\RM}  [1] {\ensuremath{\textrm{#1}}}
\newcommand {\SQRT}[1] {\ensuremath{\sqrt{#1}}}

\newcommand {\SUB} [2] {\MM{#1\ensuremath{_{#2}}}}
\newcommand {\SUP} [2] {\MM{#1\ensuremath{^{#2}}}}

\newcommand {\DIVe}[2] {\MM{#1\tsp{-1}/\tsp{-0.3}#2}}

\newcommand {\momentum} {\IT{p}}

\newcommand {\transverse} {\IT{T}}

\newcommand {\pT} {\SUB{\momentum}{\transverse}}

\newcommand {\cspeed} {\IT{c}}

\newcommand {\pico} {\RM{p}}

\newcommand {\centi} {\RM{c}}

\newcommand {\Giga} {\RM{G}}

\newcommand {\meter} {\RM{m}}

\newcommand {\cm} {\centi\meter}

\newcommand {\Volt} {\RM{V}}

\newcommand {\barn} {\RM{b}}

\newcommand {\pb} {\pico\barn}

\newcommand {\pbinv} {\SUP{\pb}{-\tsp{-0.4}1}}

\newcommand {\eV} {\tsp{0.1}\RM{e}\tsp{-0.3}\Volt}
\newcommand {\GeV} {\Giga\eV}
\newcommand {\GeVc} {\DIVe{\GeV}{\cspeed}}

\newcommand {\GeVcc} {\DIVe{\GeV}{\SUP{\cspeed}{2}}}

\newcommand {\Hz} {\RM{Hz}}

\newcommand {\zcoord} {\IT{z}}

\newcommand {\zvertex} {\SUB{\zcoord}{\RM{vertex}}}
\newcommand {\BR} {\RM{BR}}

\newcommand {\LESS} {\ensuremath{<}}

\newcommand {\LESSAPPROX} {\ensuremath{\lesssim}}
\newcommand {\GREATER} {\ensuremath{>}}

\newcommand {\DELTA} {\ensuremath{\Delta}}
\newcommand {\TIMES} {\ensuremath{\times}}

\newcommand {\svar} {\IT{s}}

\newcommand {\sqrts} {\SQRT{\tsp{-0.5}\svar\tsp{0.5}}}

\newcommand {\electron} {\IT{e}}
\newcommand {\proton} {\IT{p}}

\newcommand {\antiproton} {\ensuremath{\bar\proton}}

\newcommand {\program} [1] {\textsc{#1}}

\newcommand {\PYTHIA} {\program{pythia}}
\newcommand {\GEANT} {\program{geant}}

\newcommand {\capsword} [1] {\textsc{#1}}

\newcommand {\NLO} {\capsword{NLO}}
\newcommand {\QCD} {\capsword{QCD}}

\newcommand {\CERN} {\capsword{CERN}}

\newcommand {\RHIC} {\capsword{RHIC}}

\newcommand {\STAR} {\capsword{STAR}}

\newcommand {\BEMC} {\capsword{BEMC}}
\newcommand {\EEMC} {\capsword{EEMC}}

\newcommand {\TPC} {\capsword{TPC}}

\newcommand {\Wvar} {\IT{W}}
\newcommand {\Wpl} {\SUP{\Wvar}{+}}
\newcommand {\Wmi} {\SUP{\Wvar}{-}}
\newcommand {\WpmD} {\SUP{\Wvar}{+(-)}}
\newcommand {\epmD} {\SUP{\electron}{+(-)}}
\newcommand {\Wpm} {\SUP{\Wvar}{\pm}}
\newcommand {\epm} {\SUP{\electron}{\pm}}
\newcommand {\Zgam} {\ensuremath{\IT{Z}/\gamma^*}}

\newcommand {\pp} {\proton\proton}
\newcommand {\ppbar} {\proton\antiproton}

\newcommand {\Wtoenu} {\ensuremath{\Wvar \to e \nu}}
\newcommand {\Wptoenu} {\ensuremath{\Wpl \to e^+ \, \nu_e}}
\newcommand {\Wmtoenu} {\ensuremath{\Wmi \to e^- \, \bar{\nu}_e}}
\newcommand {\WpmDtoenu}
   {\ensuremath{\WpmD \to \epmD + \nu_e (\bar{\nu}_e)}}
\newcommand {\Wpmtoenu}{\ensuremath{\Wpm \to \epm + \nu_e}}
\newcommand {\Zgtoee} {\ensuremath{\Zgam \to e^+e^-}}

\newcommand {\EeT} {\ensuremath{E^\electron_T}}

\newcommand {\invpb} {\SUP{\pb}{-1}}
\newcommand {\NNLO} {\capsword{NNLO}}
\newcommand {\PDF} {\capsword{PDF}}

\newcommand {\Zvar} {\IT{Z}}
\newcommand {\ET}  {\ensuremath{E_T}}
\newcommand {\SIM} {\ensuremath{\sim}}
\newcommand {\Zee}  {\ensuremath{Z \to e^+e^-}}
\newcommand {\Wtaunu} {\ensuremath{W \to \tau \nu}}
\newcommand {\FEWZ} {\program{fewz}}
\newcommand {\RHICBOS} {\program{rhicbos}}
\newcommand {\mee}  {\ensuremath{m_{e^+e^-}}}

\begin{document}


\title{\texorpdfstring{Measurement of the $\bm{W \to e \nu}$ and $\bm{Z/\gamma^* \to e^+e^-}$ Production Cross Sections \\ at Mid-rapidity in Proton-Proton Collisions at $\bm{\sqrt{s}}$ = 500~GeV}{Title}}

\affiliation{Argonne National Laboratory, Argonne, Illinois 60439, USA}
\affiliation{Brookhaven National Laboratory, Upton, New York 11973, USA}
\affiliation{University of California, Berkeley, California 94720, USA}
\affiliation{University of California, Davis, California 95616, USA}
\affiliation{University of California, Los Angeles, California 90095, USA}
\affiliation{Universidade Estadual de Campinas, Sao Paulo, Brazil}
\affiliation{Central China Normal University (HZNU), Wuhan 430079, China}
\affiliation{University of Illinois at Chicago, Chicago, Illinois 60607, USA}
\affiliation{Creighton University, Omaha, Nebraska 68178, USA}
\affiliation{Czech Technical University in Prague, FNSPE, Prague, 115 19, Czech Republic}
\affiliation{Nuclear Physics Institute AS CR, 250 68 \v{R}e\v{z}/Prague, Czech Republic}
\affiliation{University of Frankfurt, Frankfurt, Germany}
\affiliation{Institute of Physics, Bhubaneswar 751005, India}
\affiliation{Indian Institute of Technology, Mumbai, India}
\affiliation{Indiana University, Bloomington, Indiana 47408, USA}
\affiliation{Alikhanov Institute for Theoretical and Experimental Physics, Moscow, Russia}
\affiliation{University of Jammu, Jammu 180001, India}
\affiliation{Joint Institute for Nuclear Research, Dubna, 141 980, Russia}
\affiliation{Kent State University, Kent, Ohio 44242, USA}
\affiliation{University of Kentucky, Lexington, Kentucky, 40506-0055, USA}
\affiliation{Institute of Modern Physics, Lanzhou, China}
\affiliation{Lawrence Berkeley National Laboratory, Berkeley, California 94720, USA}
\affiliation{Massachusetts Institute of Technology, Cambridge, MA 02139-4307, USA}
\affiliation{Max-Planck-Institut f\"ur Physik, Munich, Germany}
\affiliation{Michigan State University, East Lansing, Michigan 48824, USA}
\affiliation{Moscow Engineering Physics Institute, Moscow Russia}
\affiliation{Ohio State University, Columbus, Ohio 43210, USA}
\affiliation{Old Dominion University, Norfolk, VA, 23529, USA}
\affiliation{Panjab University, Chandigarh 160014, India}
\affiliation{Pennsylvania State University, University Park, Pennsylvania 16802, USA}
\affiliation{Institute of High Energy Physics, Protvino, Russia}
\affiliation{Purdue University, West Lafayette, Indiana 47907, USA}
\affiliation{Pusan National University, Pusan, Republic of Korea}
\affiliation{University of Rajasthan, Jaipur 302004, India}
\affiliation{Rice University, Houston, Texas 77251, USA}
\affiliation{Universidade de Sao Paulo, Sao Paulo, Brazil}
\affiliation{University of Science \& Technology of China, Hefei 230026, China}
\affiliation{Shandong University, Jinan, Shandong 250100, China}
\affiliation{Shanghai Institute of Applied Physics, Shanghai 201800, China}
\affiliation{SUBATECH, Nantes, France}
\affiliation{Texas A\&M University, College Station, Texas 77843, USA}
\affiliation{University of Texas, Austin, Texas 78712, USA}
\affiliation{University of Houston, Houston, TX, 77204, USA}
\affiliation{Tsinghua University, Beijing 100084, China}
\affiliation{United States Naval Academy, Annapolis, MD 21402, USA}
\affiliation{Valparaiso University, Valparaiso, Indiana 46383, USA}
\affiliation{Variable Energy Cyclotron Centre, Kolkata 700064, India}
\affiliation{Warsaw University of Technology, Warsaw, Poland}
\affiliation{University of Washington, Seattle, Washington 98195, USA}
\affiliation{Wayne State University, Detroit, Michigan 48201, USA}
\affiliation{Yale University, New Haven, Connecticut 06520, USA}
\affiliation{University of Zagreb, Zagreb, HR-10002, Croatia}

\author{G.~Agakishiev}\affiliation{Joint Institute for Nuclear Research, Dubna, 141 980, Russia}
\author{M.~M.~Aggarwal}\affiliation{Panjab University, Chandigarh 160014, India}
\author{Z.~Ahammed}\affiliation{Variable Energy Cyclotron Centre, Kolkata 700064, India}
\author{A.~V.~Alakhverdyants}\affiliation{Joint Institute for Nuclear Research, Dubna, 141 980, Russia}
\author{I.~Alekseev}\affiliation{Alikhanov Institute for Theoretical and Experimental Physics, Moscow, Russia}
\author{J.~Alford}\affiliation{Kent State University, Kent, Ohio 44242, USA}
\author{B.~D.~Anderson}\affiliation{Kent State University, Kent, Ohio 44242, USA}
\author{C.~D.~Anson}\affiliation{Ohio State University, Columbus, Ohio 43210, USA}
\author{D.~Arkhipkin}\affiliation{Brookhaven National Laboratory, Upton, New York 11973, USA}
\author{G.~S.~Averichev}\affiliation{Joint Institute for Nuclear Research, Dubna, 141 980, Russia}
% ...

\collaboration{STAR Collaboration}\noaffiliation

\begin{abstract}
We report measurements of the charge-separated \WpmDtoenu\ and \Zgtoee\ production cross sections at mid-rapidity in proton-proton collisions at \sqrts\ = 500~GeV.  These results are based on 13.2~\invpb\ of data recorded in 2009 by the STAR detector at RHIC.  Production cross sections for \Wvar\ bosons that decay via the $e \nu$ channel were measured to be $\sigma(pp \to \Wpl X) \cdot \BR(\Wptoenu)$ = 117.3 $\pm$ 5.9(stat) $\pm$ 6.2(syst) $\pm$ 15.2(lumi)~pb, and $\sigma(pp \to \Wmi X) \cdot \BR(\Wmtoenu)$ = 43.3 $\pm$ 4.6(stat) $\pm$ 3.4(syst) $\pm$ 5.6(lumi)~pb.  For \Zgam\ production, $\sigma(pp \to \Zgam \, X) \cdot \BR(\Zgtoee)$ = 7.7 $\pm$ 2.1(stat) $^{+0.5}_{-0.9}$(syst) $\pm$ 1.0(lumi)~pb for di-lepton invariant masses \mee\ between 70 and 110~\GeVcc.  First measurements of the \Wvar\ cross section ratio, $\sigma(pp \to \Wpl X) / \sigma(pp \to \Wmi X)$, at \sqrts\ = 500~GeV are also reported.  Theoretical predictions, calculated using recent parton distribution functions, are found to agree with the measured cross sections.
\vspace{0.6cm} 
\end{abstract}




\maketitle 


\section{\label{sec:intro}Introduction}

Studies of inclusive \Wvar\ and \Zgam\ boson production in proton-proton collisions provide valuable information, both to test the Standard Model of particle physics and to advance our understanding of the proton's substructure.  Measurements of the production cross sections $\sigma(pp \to \WpmD X) \cdot \BR(\WpmDtoenu)$ and $\sigma(pp \to \Zgam \, X) \cdot \BR(\Zgtoee)$ can be compared to theoretical calculations that involve the weak couplings between intermediate vector bosons and quarks, and which must account for higher-order terms in perturbative \QCD.  Such calculations also rely on models of the parton distribution functions ($\PDF$s) for the quarks and, in $pp$ collisions, for the antiquark `sea.'

Until recently, most measurements of \Wvar\ and \Zgam\ production in hadronic interactions have been confined to experiments using proton-antiproton collisions.  First results were obtained by the UA1 \cite{Albajar:1987yz,Albajar:1988ka} and UA2 \cite{Alitti:1990gj,Alitti:1991dm} collaborations at \sqrts\ = 630~GeV at the \CERN\ $\rm{Sp \bar{p} S}$ facility, followed by the CDF \cite{Abe:1995bm,Abulencia:2005ix} and D0 \cite{Abachi:1995xc,Abbott:1999tt} \ppbar\ measurements at the Fermilab Tevatron, at \sqrts\ = 1.8 and 1.96~TeV.  It is only in the last few years that \pp\ colliders have reached sufficient center of mass energies for comparable studies, at \sqrts\ = 500~GeV by the \STAR\ \cite{Aggarwal:2010vc} and PHENIX \cite{Adare:2010xa} collaborations at the Relativistic Heavy Ion Collider (RHIC), and most recently by the LHC experiments ATLAS \cite{Aad:2010yt} and CMS \cite{Khachatryan:2010xn,Chatrchyan:2011nx} at \sqrts\ = 7~TeV.  

RHIC is unique in its capability to collide high energy polarized proton beams, and the observation of \Wvar\ production in these polarized proton collisions provides a new means to explore the spin-flavor structure of proton sea quark distributions.  First measurements of the parity-violating longitudinal single-spin asymmetry for \Wpm\ decay leptons have also been reported by the \STAR\ \cite{Aggarwal:2010vc} and PHENIX \cite{Adare:2010xa} collaborations and are in good agreement with predictions from \NLO\ and resummed calculations \cite{deFlorian:2010aa,Nadolsky:2003ga}.

At hadron colliders, the leading process in \WpmD\ production is $u+\bar{d} \, (d+\bar{u})$ fusion.  This suggests that while the \Wpl\ and \Wmi\ production cross sections should be close to equal in $p\bar{p}$ collisions, they can be expected to differ in $pp$ measurements due to differences in the $u$ and $d$ quark and antiquark distributions within the proton.  The $\PDF$s that characterize the valence $u$ and $d$ quarks of the proton (or $\bar{u}$ and $\bar{d}$ in the antiproton) are well determined from decades of high precision, deep-inelastic lepton scattering experiments (see, for example, Ref.~\cite{Martin:2002aw}).  Comparable distributions for the antiquarks within the proton sea, however, are much more weakly constrained. Interest in these poorly-known antiquark \PDF's has also increased over the last few years, due to results from Drell-Yan experiments \cite{Baldit:1994jk,Towell:2001nh} which find evidence for a much larger $\bar{d}/\bar{u}$ flavor asymmetry in the nucleon than had been anticipated, especially at momentum fractions near and above $x \sim 0.2$.  Detailed measurements of \Wpm\ and \Zgam\ production in proton-proton collisions will provide new and complementary information about this flavor asymmetry in the sea, from different reactions and at very different momentum scales.

This paper describes the first measurement of the \Wpl, \Wmi, and \Zgam\ boson production cross sections in proton-proton collisions at $\sqrt{s}=500$ GeV by the STAR collaboration at RHIC.  The cross sections are derived from studies of the charge-separated \WpmDtoenu\ and \Zgtoee\ decay channels for outgoing leptons near mid-rapidity ($|\eta_e| < 1$), and are based on 13.2~\invpb\ of data recorded during the 2009 run.  In addition to the individual cross sections, a first measurement of the \Wpl/\Wmi\ cross section ratio at \sqrts\ = 500~GeV is also presented.

The paper is organized as follows.  Section~\ref{sec:detector} provides a brief overview of the STAR detector, focusing on the subsystems used in this analysis.  Section~\ref{sec:data} describes the data and simulation samples analyzed, Sec.~\ref{sec:signal} details the extraction of the \Wvar\ and \Zgam\ signal spectra, and Sec.~\ref{sec:background} explains the estimation and subtraction of the background from the signal spectra.  
Finally, we discuss the calculation of the \Wvar\ and \Zgam\ production cross sections in Sec.~\ref{sec:xsec} and the \Wpl/\Wmi\ cross section ratio in Sec.~\ref{sec:ratio}, and compare these results to several theoretical calculations.  Some of the data analysis methods employed here have been described briefly in Ref.~\cite{Aggarwal:2010vc}, and are discussed in more detail in this paper which incorporates a slightly larger data sample as well as improved detector calibrations with respect to the previous publication.

\section{\label{sec:detector}The STAR Detector}

\begin{figure}[!ht]
  \includegraphics[width=1.0\columnwidth]{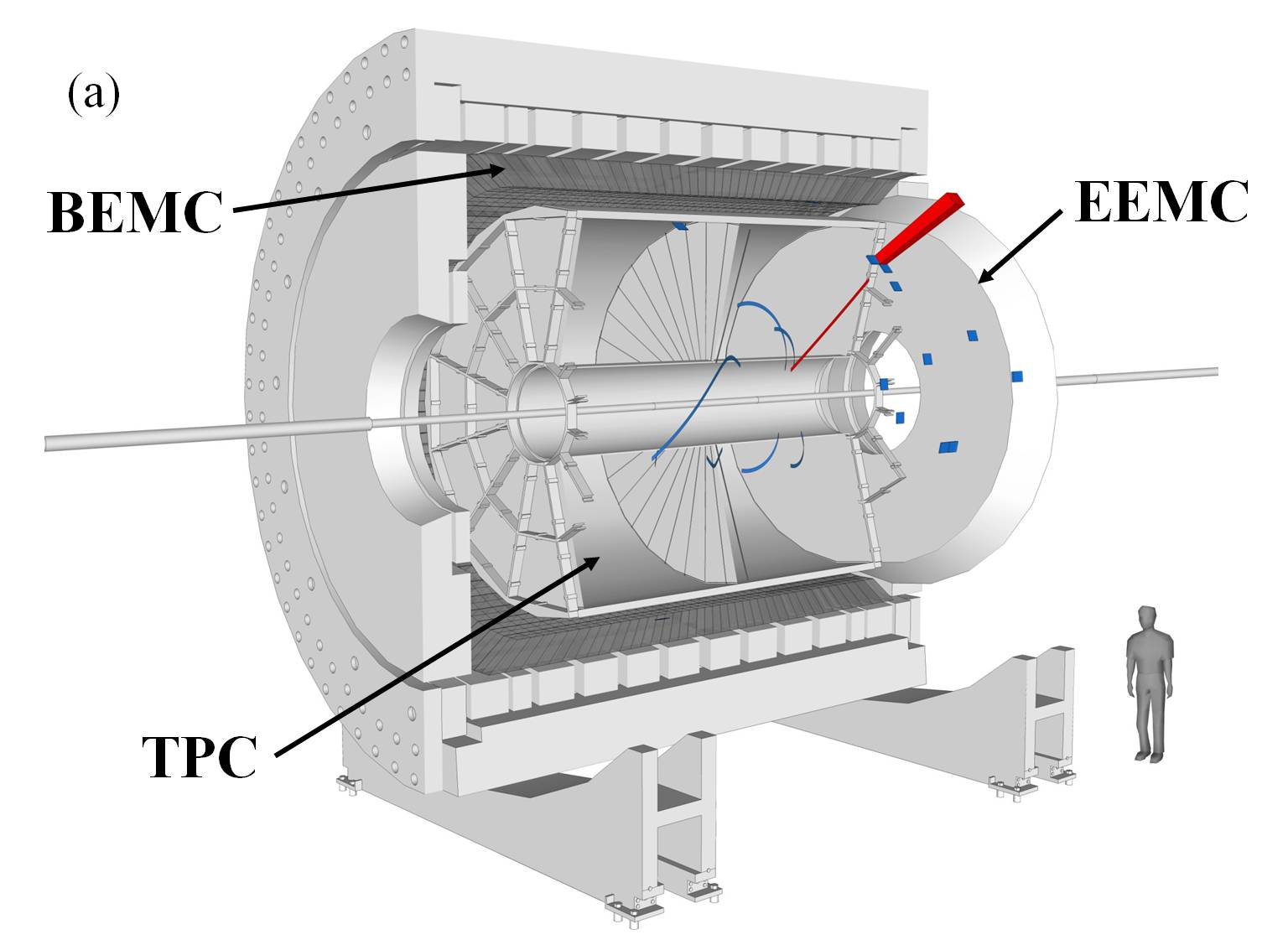}
  \includegraphics[width=1.0\columnwidth]{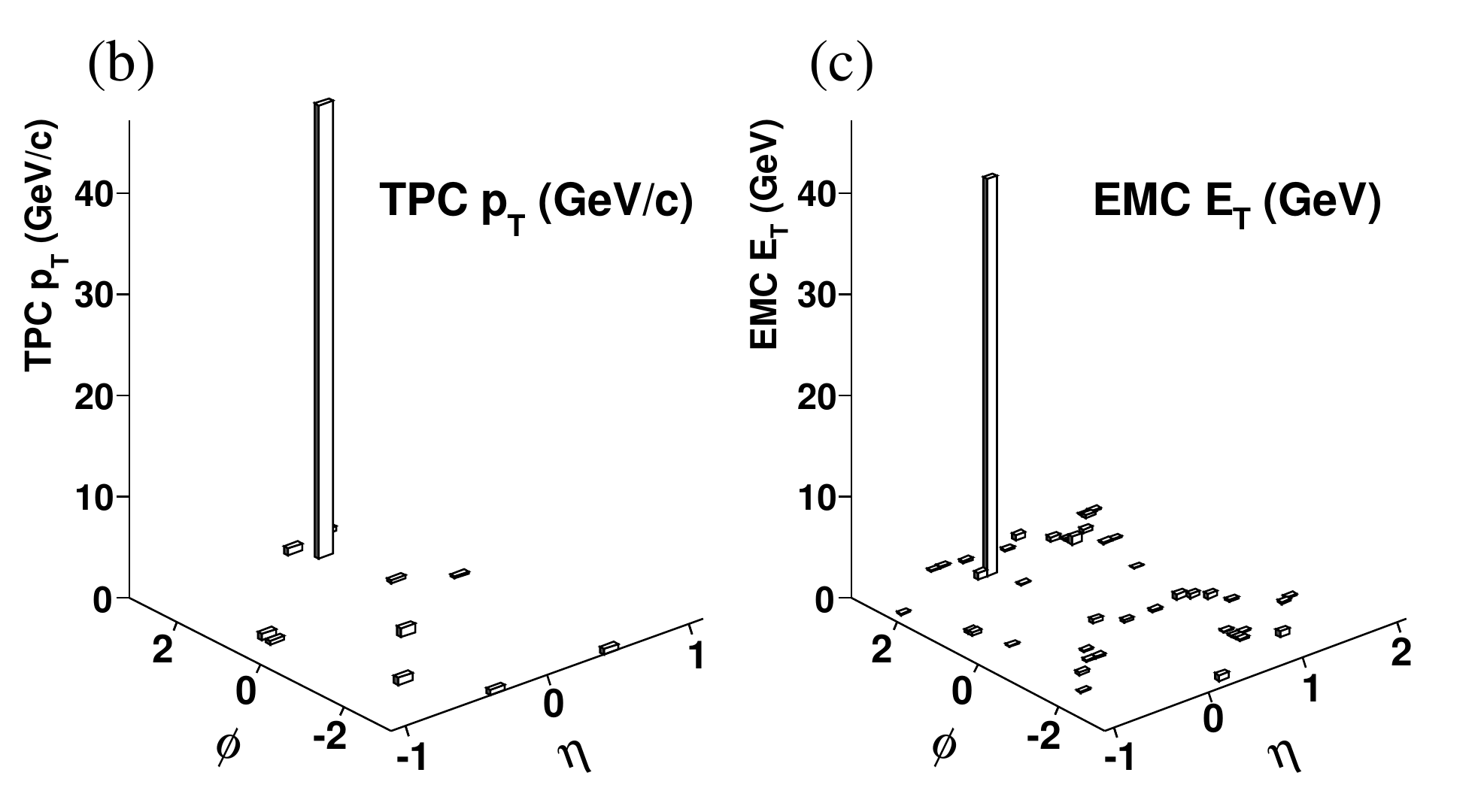}
  \caption{(Color online) (a) \Wvar\ candidate event display embedded in a schematic of the \STAR\ detector.  Descriptions of the subsystems relevant for this analysis are given in Sec.~\ref{sec:detector}.  (b) TPC \pT\ and (c) BEMC and EEMC \ET\ distributions in $\eta$ and $\phi$ for the same \Wvar\ candidate event as shown in (a). }
  \label{fig:STAR}
\end{figure}

The \STAR\ detector (Solenoidal Tracker at RHIC) \cite{Ackermann:2002ad}, shown schematically in Fig.~\ref{fig:STAR}, is a large acceptance, multipurpose detector designed primarily for measurements of hadronic and electromagnetic particle production in high-energy heavy ion and polarized proton-proton collisions. \STAR\ is comprised of many separate subsystems, each with specific capabilities; only those subsystems most relevant for the present analysis will be mentioned below.

The heart of \STAR\ is a large Time Projection Chamber (\TPC) \cite{Anderson:2003ur} which is situated within a highly uniform, 0.5~T solenoidal magnetic field.  The \TPC\ provides charged particle tracking, particle identification (via ionization energy loss, $dE/dx$), and precision momentum measurements over the range $|\eta| < 1.3$ and with full $2\pi$ azimuthal coverage.  Although the \pT\ resolution of the \TPC\ deteriorates with increasing \pT, the spacial accuracy of tracks reconstructed between the inner and outer radius of the $\TPC$, located at 50 and 200~cm respectively, remains accurate up to $\SIM$1-2~mm in Cartesian space.  In this analysis, \TPC\ tracks were used in identifying the high-\pT\ decay lepton (\epm) candidates, determining candidate charge signs, reducing contamination from the significant QCD background (see Sec.~\ref{sec:signal}), and reconstructing the interaction vertex for the events of interest.

Surrounding the \TPC\ radially is the Barrel Electromagnetic Calorimeter (\BEMC) \cite{Beddo:2002zx}, a high granularity lead/scintillator-based sampling calorimeter.  This detector is used to measure the energy deposited by energetic photons and electrons with pseudorapidities $|\eta| < 1.0$ over the full azimuth.  The \BEMC\ is segmented into 4800 optically isolated projective towers, each of which subtends 0.05~rad in azimuth ($\phi$) and 0.05 units in $\eta$, and is roughly 20 radiation lengths deep. Based on cosmic ray and test beam data, the nominal energy resolution of the barrel calorimeter is calculated to be $\delta E/E = 14\%/\sqrt{E\mathrm{(GeV)}} \oplus 1.5\%$ \cite{Beddo:2002zx}.  The \BEMC\ was used to measure the \epm\ candidate energy, and to aid in background reduction.  By identifying events with large, highly localized, electromagnetic energy deposition, the \BEMC\ also provided our first-level trigger signal for leptonic \Wvar\ and \Zvar\ decays.

Located at one end of the \STAR\ \TPC, directly in front of the magnetic field return poletip, is the Endcap Electromagnetic Calorimeter (\EEMC) \cite{Allgower:2002zy}, which provides electromagnetic energy measurement over the range $1.09 < \eta < 2$ and 2$\pi$ in azimuth.  The \EEMC\ is similar in design to the \BEMC: a lead/scintillator sampling calorimeter, finely segmented in $\eta$ and $\phi$ into 720 towers with projective geometries, though it is approximately 3-4 radiation lengths thicker than the \BEMC\ due to its more forward position.  In the work presented here, the \EEMC\ was used only as part of the background reduction via isolation and missing energy conditions discussed in Sec.~\ref{sec:signal}.

\section{\label{sec:data}Data and Simulation Samples}

Candidate events were selected online using a two-level trigger requirement in the $\BEMC$.  The hardware level-0 trigger accepted events containing a tower with a transverse energy, $\ET$, greater than 7.3~\GeV.  A dedicated software trigger algorithm then selected events by constructing 2$\TIMES$2 clusters of towers, and requiring that at least one cluster consist of a seed tower with $\ET \GREATER$~5~$\GeV$ and a cluster sum $\ET~\GREATER$~13~$\GeV$.  During the 2009 run $1.2\TIMES10^6$ events were recorded satisfying these trigger conditions.

\begin{figure}[!ht]
  \includegraphics[width=1.0\columnwidth]{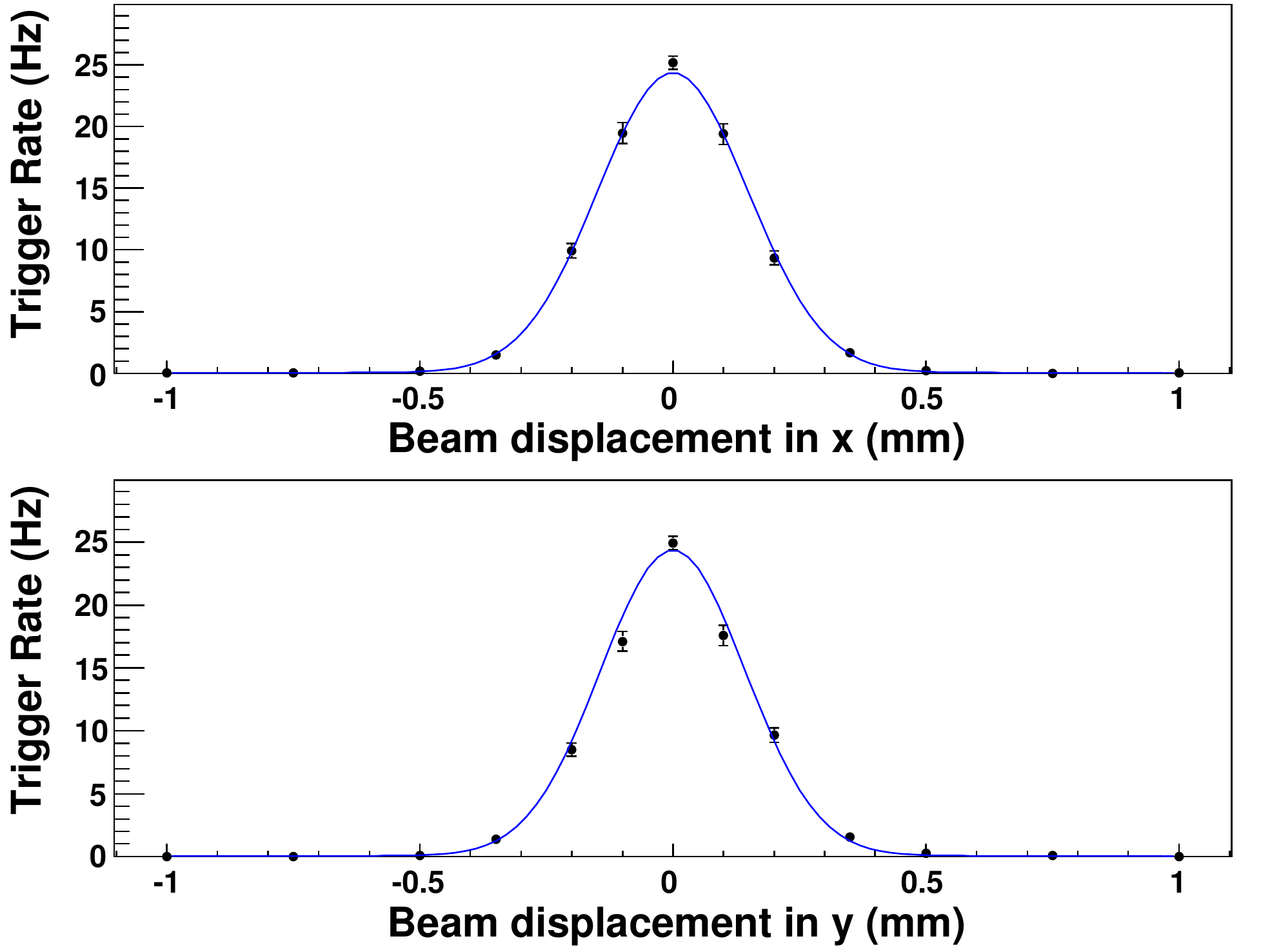}
  \caption{(Color online) Trigger rate as a function of vernier scan beam displacement in the $x$ and $y$ directions.  The transverse beam widths ($\sigma_x$ and $\sigma_y$) and maximum trigger rate ($\RM{R}^{\RM{max}}_{\RM{ver}}$) were extracted from the fit, which is superimposed. }  
  \label{Fig:vernier}
\end{figure}

The integrated luminosity of the data sample was determined using the Vernier Scan technique \cite{SvdeMeer1968}.  The transverse widths ($\sigma_x$ and $\sigma_y$) of the beam overlap region are determined by measuring the trigger rate as the beams are swept through each other in the transverse plane.  The intensity of each beam is determined during a scan by the Wall Current Monitors (WCM) \cite{WCM}.  With the assumption of Gaussian beams, the instantaneous luminosity can be written as
\begin{equation}
\mathcal{L}=\frac{f_{rev}K}{2\pi\sigma_x\sigma_y}
\end{equation}
where $f_{rev}$ is the revolution frequency and $K=\sum N^a_iN^b_i$ is the product of the bunch intensities ($N_i$) of the two beams ($a$,$b$) summed over all bunches.  The dedicated trigger used in the Vernier Scan, and also to monitor the luminosity in this analysis, is the level-0 hardware trigger, described above, with a coincidence away-side $\ET$ requirement imposed offline to reduce non-collision background.  The cross section for this trigger can be written as $\sigma_{\RM{ver}}=\RM{R}^{\RM{max}}_{\RM{ver}}/\mathcal{L}$, where $\RM{R}^{\RM{max}}_{\RM{ver}}$ is the maximum trigger rate while the beams are fully overlapping.  The value measured for this work was $\sigma_{\RM{ver}}$ = 434 $\pm$ 8(stat) $\pm$ 56(syst)~nb.  Figure \ref{Fig:vernier} shows an example of the trigger rate as a function of the $x$ and $y$ beam displacements during one of the vernier scans, which was fit to extract the transverse beam widths and maximum trigger rate.  The fit function used was a Gaussian in $x$ and $y$ combined with a constant term to account for remaining non-collision background.  The largest contribution to the $\sigma_{\RM{ver}}$ systematic uncertainty was attributed to possible non-Gaussian components of the beam profile (10\%), with smaller contributions coming from possible \BEMC\ gain drift (5\%), and uncertainties in the bunch intensity measurements (4\%).  This value for $\sigma_{\RM{ver}}$ was used to normalize the total number of events which satisfy this trigger condition, resulting in an integrated luminosity for the data sample of $L$ = $\int\mathcal{L}~dt$ = 13.2 $\pm$ 0.2(stat) $\pm$ 1.7(syst)~\invpb.

Monte-Carlo (MC) simulation samples were generated in order to determine detector efficiencies, estimate background contributions from electroweak processes, and compare various predicted observables to data.  Signal samples for both the \Wtoenu\ and \Zgtoee\ channels were generated, along with a \Wtaunu\ sample which is an expected background in the \Wvar\ analysis due to the $\tau$'s leptonic decay.  All the samples were produced using the \PYTHIA\ 6.422 \cite{Sjostrand:2006za} event generator and a \GEANT\ \cite{Brun:1978fy} model of the \STAR\ detector response.  The same reconstruction and analysis algorithm was used for both the data and MC samples, and each MC sample was normalized to the integrated luminosity of the data unless otherwise stated.

Due to the high luminosity of the \pp\ collision environment at \sqrts\ = 500~\GeV\ at \STAR, a significant number of pile-up tracks are present in the \TPC\ at any given time.  The pile-up tracks are the result of either another collision from the same bunch crossing as the triggered event, or a collision that occurred in an earlier or later bunch crossing.  Note that the bunch crossing period at RHIC is about 107~ns, while it can take up to $\sim$38~$\mu$s for track ionization to drift through the \TPC.  In the simulation, these pile-up tracks are accounted for by embedding the full \GEANT\ detector response of the simulated event into a zero-bias triggered event before reconstruction.  The zero-bias events are selected randomly during nominal beam crossings at a rate of $\LESSAPPROX 1 \Hz$ with no detector requirements, resulting in a good representation of the pile-up contained in the \TPC\ for \BEMC\ triggered collision events. 

\section{\label{sec:signal}\texorpdfstring{$\bm{\Wvar}$ and $\bm{\Zgam}$ Signal Reconstruction}{W and Z/gamma* Signal Reconstruction}}

This section details the identification and reconstruction of $\Wvar$ and $\Zgam$ candidate events, as well as the reduction of the large QCD background.  This reduction is achieved through a number of cuts designed to take advantage of the kinematical and topological differences between electroweak and QCD processes.  ``\Zgam" will be used interchangeably with ``\Zvar" for the remainder of this paper.

Candidate events were selected from the sample of \BEMC\ triggered events described in Sec.~\ref{sec:data} by requiring a reconstructed primary vertex.  A primary vertex is one reconstructed from either a single \TPC\ track with $\pT \GREATER 10~\GeVc$ or multiple tracks originating from the same location along the beamline.  Each track considered in vertex reconstruction is assigned an increased weight if it either points to a region of energy deposition in the calorimeters, or if it uses hit points from both sides of the \TPC\ central membrane.  Tracks satisfying either of these two conditions are likely to be from the triggered collision, therefore weighting these tracks more heavily in vertex reconstruction strongly reduces the contamination from pile-up tracks.  The distribution of primary vertices along the beam direction is approximately Gaussian with an RMS width of 52 cm.  Events of interest were required to have a $|\zvertex|\LESS100$ cm, where $\zvertex$ is the distance along the beam direction of the primary vertex from the nominal collision point at the center of the STAR interaction region.
 
\subsection{\label{subsec:epm_isolation}\texorpdfstring{Identification of High-$\bm{\ET}$ Isolated Electrons and Positrons}{Identification of High-ET Isolated Electrons and Positrons}}

A candidate electron or positron track is defined to be a \TPC\ track with $\pT \GREATER 10~\GeVc$ that is associated with a primary vertex satisfying the criteria described above.  Candidate tracks were also required to have:
\begin{itemize}\addtolength{\itemsep}{-0.5\baselineskip} 
\item a minimum of 15 \TPC\ points,
\item more than 51\% of the maximum number of \TPC\ points allowed,
\item a first \TPC\ point with radius less than 90 cm,
\item a last \TPC\ point with radius greater than 160 cm.
\end{itemize}
These requirements help to ensure that the track and its charge sign are well reconstructed, as well as reject pile-up tracks which may be mistakenly associated with a primary vertex.  Candidate \TPC\ tracks are extrapolated to the \BEMC\ to determine to which tower the track points, then the four possible 2$\TIMES$2 \BEMC\ tower clusters containing the tower pointed to by the track are formed.  The 2$\TIMES$2 cluster with the largest summed transverse energy, $\EeT$, is assigned to the $\epm$ candidate.  The candidate $\EeT$ is required to be greater than 15 $\GeV$ to be safely above the trigger turn-on region.  Also, the two dimensional distance between the energy log-weighted centroid of the tower cluster position and the extrapolated \TPC\ track position, $|\DELTA\vec{r}|$, is required to be less than 7 cm, to reject candidates where the \BEMC\ cluster may not have originated from the particle which produced the high \pT\ \TPC\ track. 

\begin{figure}[!ht]
  \includegraphics[width=1.0\columnwidth]{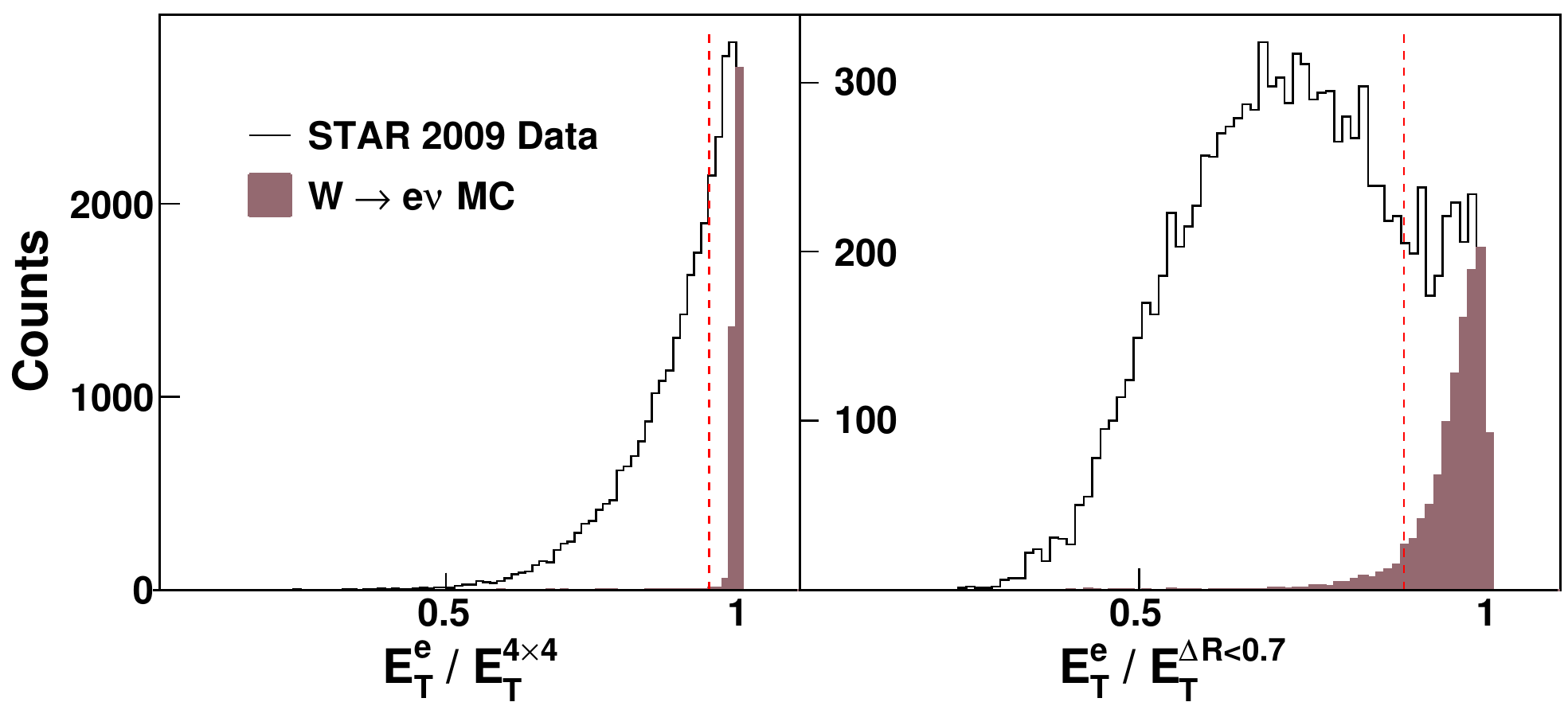}
  \caption{(Color online) Distributions of the isolation ratios $\EeT/\ET^{4\TIMES4}$ (left) and $\EeT/\ET^{\DELTA\RM{R}\LESS0.7}$ (right) used in \epm\ candidate selection.  \Wtoenu\ MC shape distributions (arbitrary normalization) are shown as filled histograms for comparison with the data distributions.  The vertical dashed lines indicate the placements of the cuts on these isolation ratios.}
  \label{Fig:isoCuts}
\end{figure}

Electrons and positrons from \Wvar\ and \Zvar\ decays should be well isolated from other particles in $\eta-\phi$ space; thus, in the next stage of the candidate selection process two isolation criteria are applied.  The first isolation cut was made by summing the \ET\ in the 4$\TIMES$4 \BEMC\ tower cluster which surrounds the $\epm$ candidate cluster, $\ET^{4\TIMES4}$, and requiring $\EeT/\ET^{4\TIMES4} \GREATER$ 0.95.  The other isolation requirement is imposed to reduce jet-like events.  The quantity $\ET^{\DELTA\RM{R}\LESS0.7}$ is defined as the sum of all \BEMC\ and \EEMC\ tower $\ET$ and \TPC\ track $\pT$ within a cone radius of $\DELTA\RM{R}=\sqrt{\DELTA\eta^2+\DELTA\phi^2}\LESS0.7$ around the candidate track, and the ratio $\EeT/\ET^{\DELTA\RM{R}\LESS0.7}$ is required to be greater than 0.88.  The \epm\ candidate track is excluded from the sum of \TPC\ track $\pT$ to avoid double-counting the candidate energy in the $\ET^{\DELTA\RM{R}\LESS0.7}$ sum.  Figure \ref{Fig:isoCuts} shows the isolation ratios described above for both data and \Wtoenu\ MC.  The placements of the cuts, shown by the dashed lines, were chosen to retain a large fraction of the signal, while significantly reducing the QCD background.  Note that differences between the isolation ratios in Fig.~\ref{Fig:isoCuts} of this paper and Fig.~1 of Ref.~\cite{Aggarwal:2010vc} are expected due to differences in the data samples used and improved calibrations.  Also, the order of the $\EeT/\ET^{4\TIMES4}$ and candidate track-cluster matching $|\DELTA\vec{r}|$ cuts were inverted in Ref.~\cite{Aggarwal:2010vc} with respect to the ordering described in this section.  

\subsection{\label{subsec:Wsignal}\texorpdfstring{$\bm{\Wvar}$ Candidate Event Selection}{W Candidate Event Selection}}

The selection of \Wtoenu\ candidate events is based on differences in the event topology between leptonic \Wvar\ decays and the QCD background or \Zvar\ events.  \Wtoenu\ events contain a nearly isolated \epm\ with a neutrino close to opposite in azimuth.  Electrons and positrons emitted near mid-rapidity from \Wvar\ decay are characterized by a large \EeT\ that is peaked near half the \Wvar\ mass (\SIM40~\GeV) with a distribution referred to as a Jacobian peak.  There is also a large missing transverse energy in \Wtoenu\ events opposite, in azimuth, to the \epm\ due to the undetected neutrino.  As a result, there is a large imbalance in the vector \pT\ sum of all reconstructed final state objects for \Wvar\ events.  In contrast, \Zee\ events and QCD hard-scattering events, such as di-jets, are characterized by a small magnitude of this vector \pT\ sum imbalance.

\begin{figure}[!ht]
  \includegraphics[width=1.0\columnwidth]{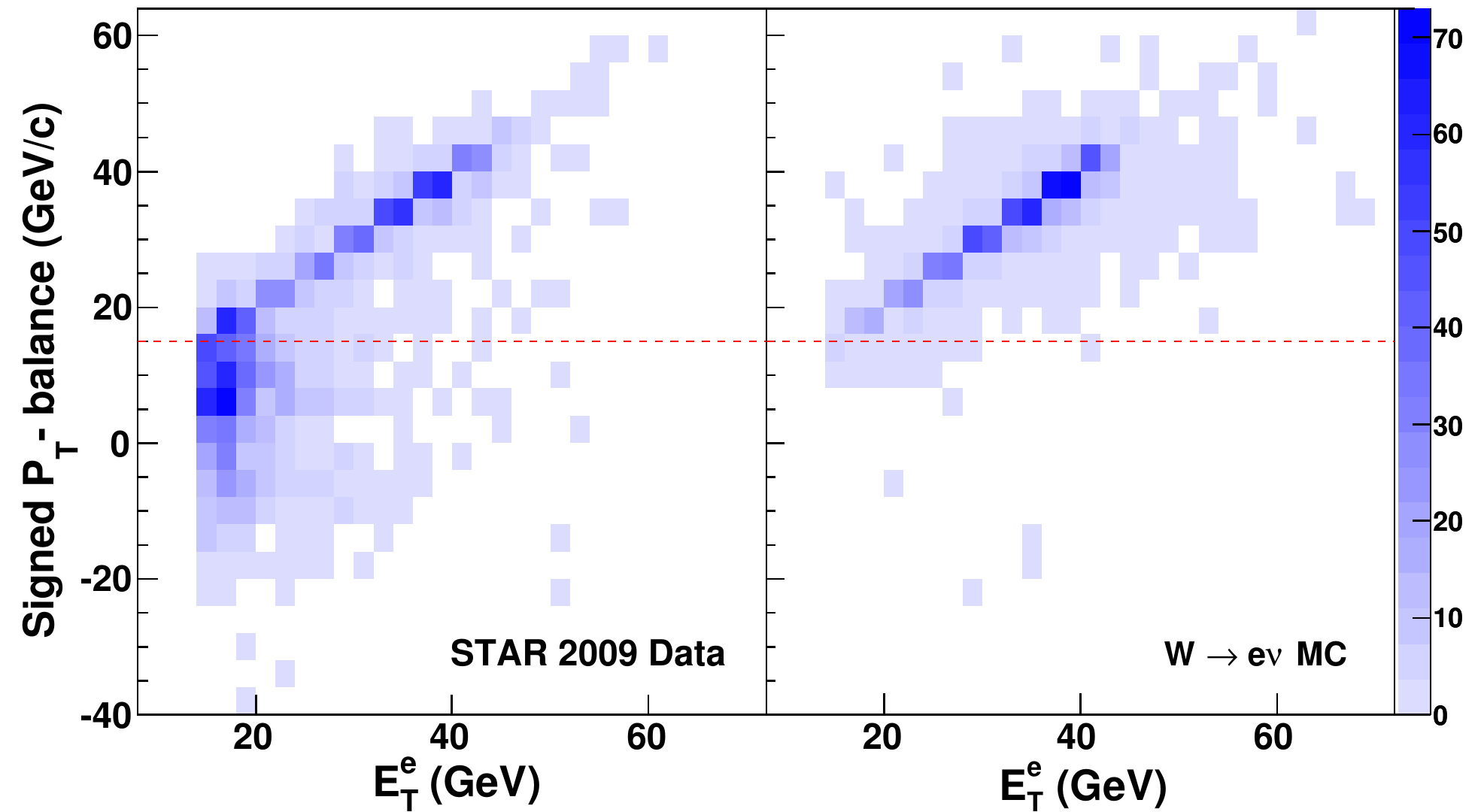}
  \caption{(Color online) Correlation of the signed $P_T$-balance variable and \EeT\ for data (left) and \Wtoenu\ MC (right).}
  \label{Fig:sPtBalcut}
\end{figure}

In order to enforce this missing energy requirement, we define the \pT\ balance vector:
\begin{equation}
  \vec{p}_{T}^{~bal} = \vec{p}_{T}^{~e} + \sum_{\DELTA\RM{R}>0.7} \vec{p}_{T}^{~jets} 
  \label{eqn:ptBal}
\end{equation}
where $\vec{p}_{T}^{~e}$ is the \epm\ candidate \pT\ vector,  which is composed of a momentum direction and a magnitude determined by the candidate TPC track and BEMC cluster, respectively.  The second term on the right of Eq.~\ref{eqn:ptBal} is the sum of the \pT\ vectors for all reconstructed jets whose thrust axes are \textit{outside} the cone radius of $\DELTA\RM{R}=0.7$ around the candidate.  Jets are reconstructed using a standard mid-point cone algorithm used in STAR jet measurements \cite{Abelev:2006uq} based on the tracks from the \TPC\, and tower energies in the \BEMC\ and \EEMC.  A scalar signed $P_T$-balance variable is then formed, defined as
\begin{equation}
  \mbox{signed }P_{T}\mbox{-balance} = \mbox{sign}\left(\vec{p}_{T}^{~e} \cdot \vec{p}_{T}^{~bal}\right) \left|\vec{p}_{T}^{~bal}\right|.
\end{equation} 
This quantity is required to be larger than 15 \GeVc\ as indicated by the dashed line in Fig.~\ref{Fig:sPtBalcut}.  Also in Fig.~\ref{Fig:sPtBalcut} one can see that in the \Wtoenu\ MC sample, the signed $P_T$-balance variable and $\EeT$ are very well correlated, as contributions to the $\vec{p}_{T}^{~bal}$ vector from reconstructed jets outside the cone of $\DELTA\RM{R}=0.7$ are generally small.  The data show a similar correlation at high \EeT, where the distribution is dominated by \Wtoenu\ events.  At low \EeT\, where contributions from QCD background events are larger, more events have a small value for the signed $P_T$-balance variable, as expected.

Background events from \Zee\ decays are further suppressed by rejecting events with an additional $e$-like 2$\TIMES$2 cluster in a reconstructed jet where $\ET^{2\TIMES2} > p^{jet}_T/2$ and the invariant mass of the two $\epm$-like clusters is within the range of 70 to 140 \GeVcc.  This reduces $\Zee$ contamination in both the \Wvar\ signal spectra and in the spectra that will be used for the data-driven QCD background, described in Sec.~\ref{subsec:Wback}.

\begin{figure}[!ht]
  \includegraphics[width=1.0\columnwidth]{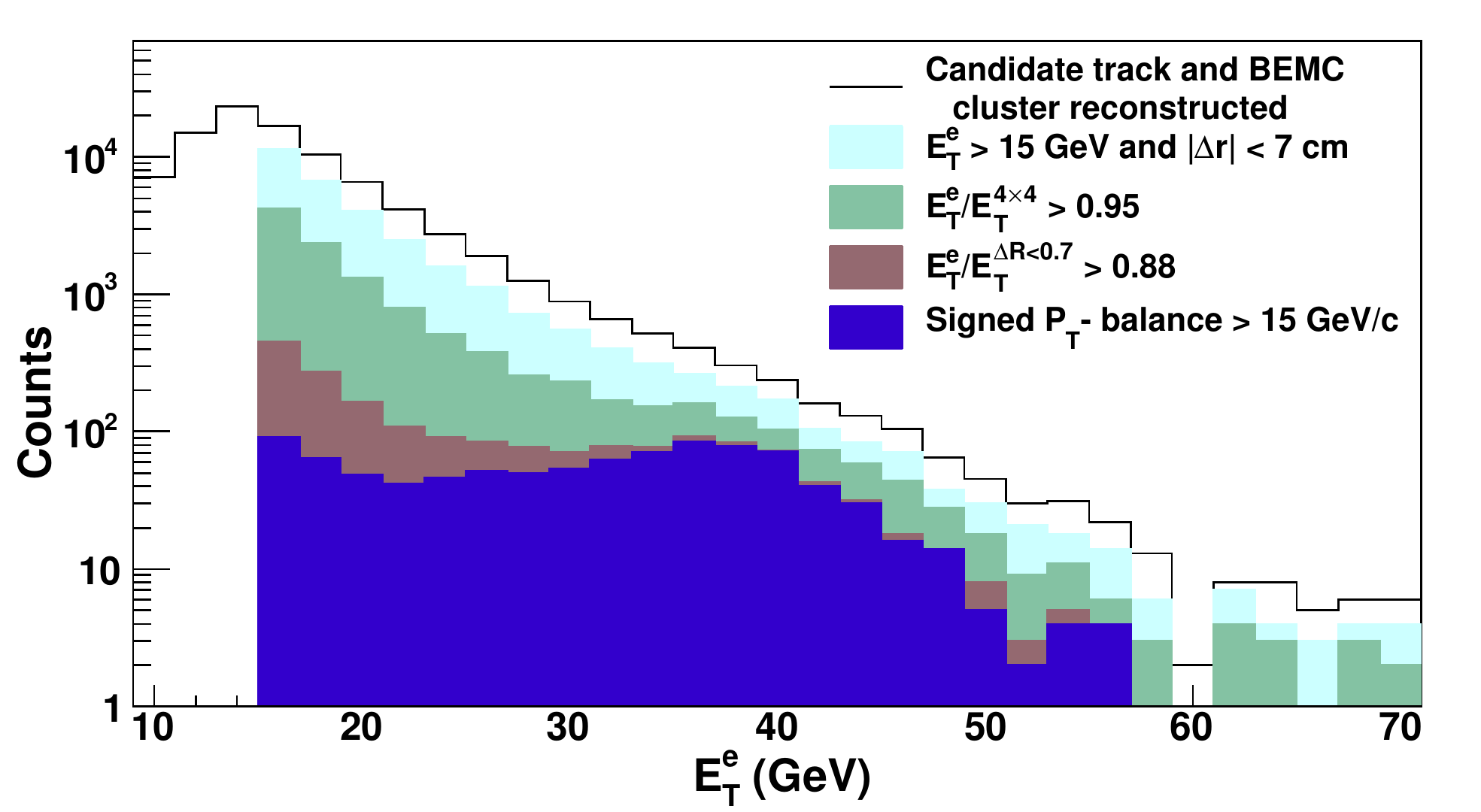}
  \caption{(Color online) Distributions of \EeT\ for \Wvar\ candidate events after sequentially applying the selection criteria described in Secs.~\ref{subsec:epm_isolation} and \ref{subsec:Wsignal}.}
  \label{Fig:wStack}
\end{figure}

The reduction in the \Wvar\ candidate yield after each of the selection criteria is shown in Fig.~\ref{Fig:wStack}.  Initially, when only a candidate \TPC\ track and \BEMC\ cluster have been reconstructed, the distribution (solid line) is dominated by QCD background, which is exponentially falling with \EeT, and there is no evidence of the Jacobian peak.  However, once the \epm\ selection, isolation and signed $P_T$-balance cuts are applied, a \Wvar\ signal can be seen above the background at $\EeT \SIM M_W/2$.

\begin{figure}[!ht]
  \includegraphics[width=1.0\columnwidth]{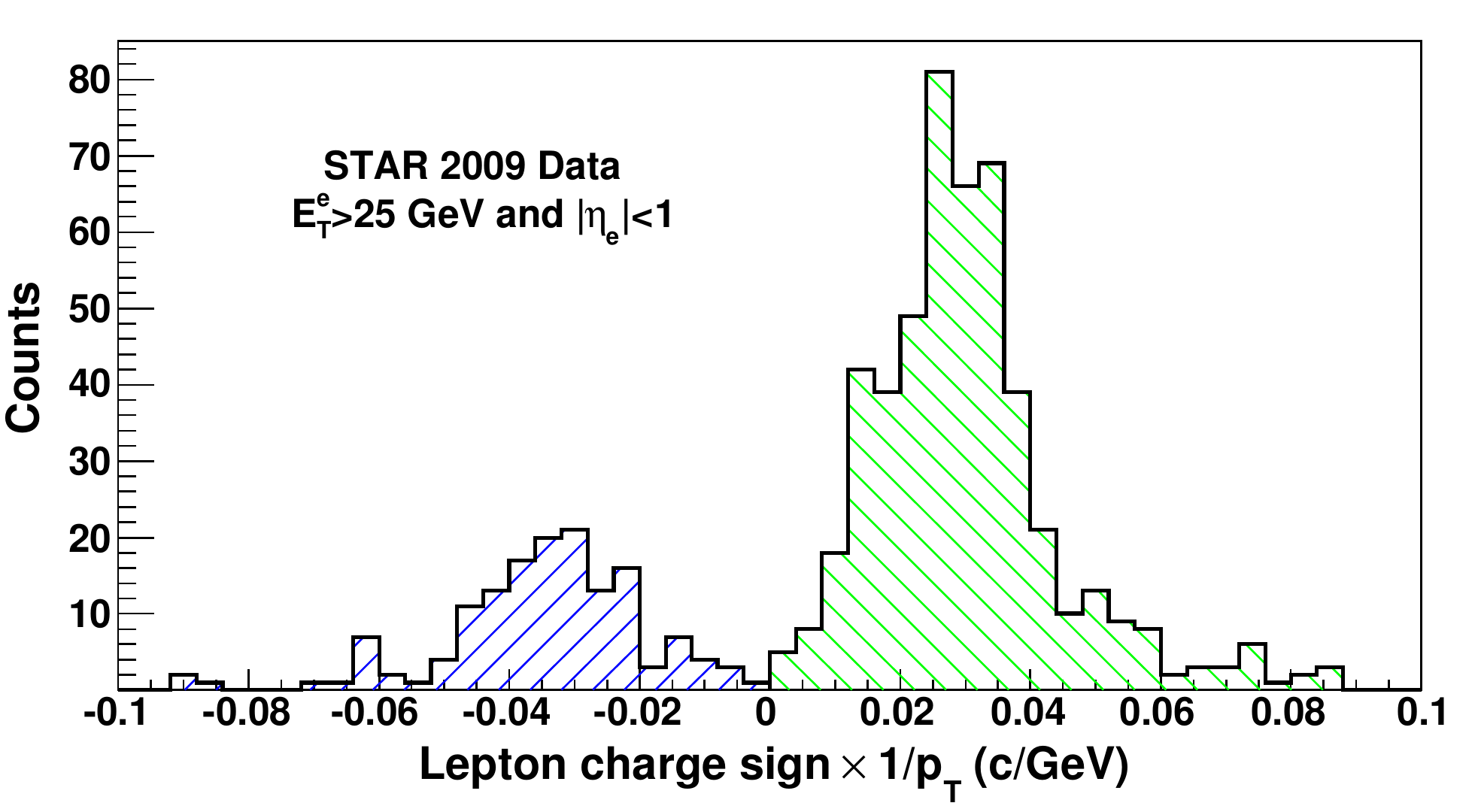}
  \caption{(Color online) Distribution of the product of the \TPC\ reconstructed charge sign and 1/$\pT$ for candidates satisfying all the \Wvar\ signal selection criteria and $\EeT \GREATER 25~\GeV$.}
  \label{Fig:chargesep}
\end{figure}

The charge sign of the \epm\ candidate is determined by the direction of curvature of the \TPC\ track in the \STAR\ magnetic field, while the magnitude of the track curvature provides a measure of 1/$\pT$.  Figure \ref{Fig:chargesep} shows the product of the reconstructed charge sign and 1/$\pT$ for the lepton candidates that satisfy all the cuts described above with $\EeT \GREATER 25~\GeV$.  Two well-separated regions are seen for the positive and negative charges, cleanly distinguishing between the $e^+$ and $e^-$ candidates. 

\subsection{\label{subsec:Zsignal}\texorpdfstring{$\bm{\Zvar}$ Candidate Event Selection}{Z Candidate Event Selection}}

\begin{figure}[!ht]
  \includegraphics[width=1.0\columnwidth]{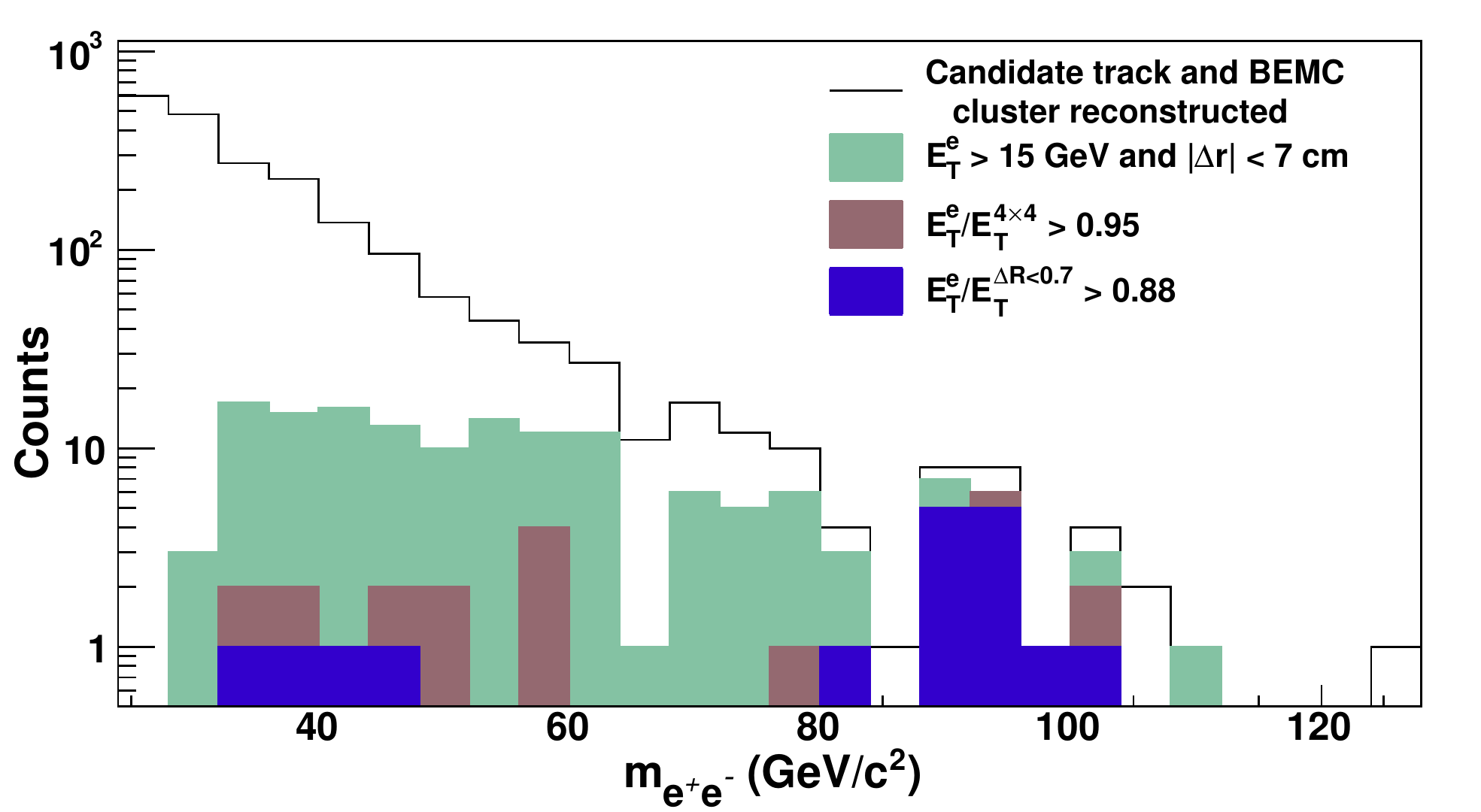}
  \caption{(Color online) Distributions of the invariant mass of pairs of oppositely charged \epm\ candidates after sequentially applying the selection criteria described in Sec.~\ref{subsec:epm_isolation} to both \epm\ candidates.}
  \label{Fig:zStack}
\end{figure}

\begin{figure}[!ht]
  \includegraphics[width=1.0\columnwidth]{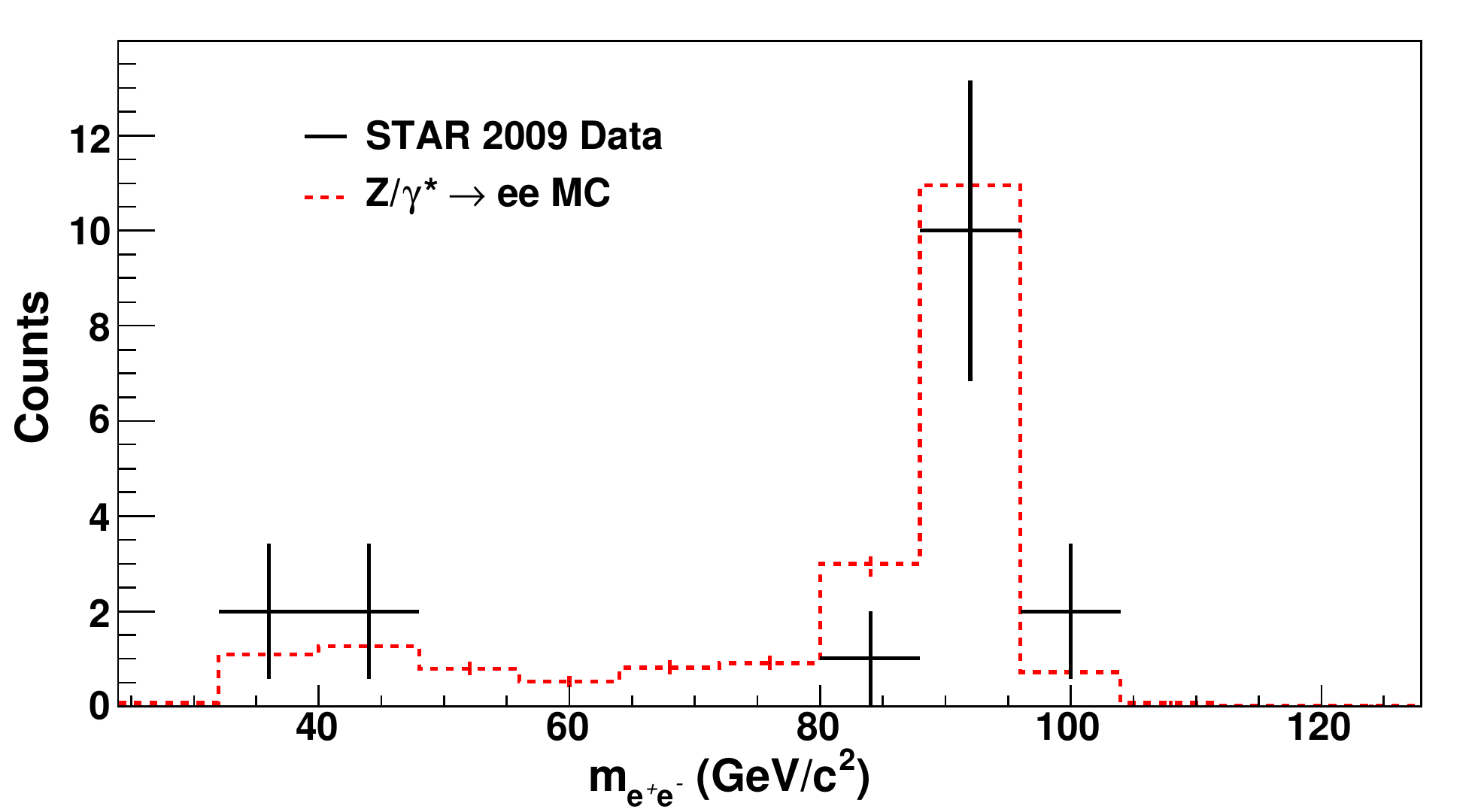}
  \caption{(Color online) Distributions of the invariant mass of \Zgtoee\ candidate events satisfying all selection criteria described in Sec.~\ref{subsec:Zsignal}.  The \Zgtoee\ MC distribution (dashed line) is shown for comparison.  Note the larger bin widths relative to Fig.~\ref{Fig:zStack}}
  \label{Fig:zDataMC}
\end{figure}

Using the isolated \epm\ sample found in Sec.~\ref{subsec:epm_isolation}, \Zee\ events were selected by requiring a pair of isolated \epm\ candidates with opposite charge signs.  The invariant mass of each $e^+e^-$ pair was reconstructed, and the resulting mass distributions are shown in Fig.~\ref{Fig:zStack} after each of the selection criteria described in Sec.~\ref{subsec:epm_isolation} has been satisfied for both the $e^+$ and $e^-$ candidates. 
After all selection cuts are applied, there is a signal near the invariant mass of the \Zvar\ and a small signal at lower invariant mass.  This is consistent with the expectations from the \Zgtoee\ MC, as shown in Fig.~\ref{Fig:zDataMC}.

\section{\label{sec:background}Background Estimation}

\subsection{\label{subsec:Wback}\texorpdfstring{$\bm{\Wvar}$ Background Estimation}{W Background Estimation}}

There are a number of background processes that can contribute to the \Wtoenu\ candidate yield.  Other electroweak processes also yield isolated electrons that can be misidentified as \Wtoenu\ events, and QCD jets can fragment in such a way that they satisfy all the \Wvar\ signal requirements.  This section describes how the contributions of these background processes to the \Wvar\ candidate yield are estimated.

\begin{figure}[!ht]
  \includegraphics[width=1.0\columnwidth]{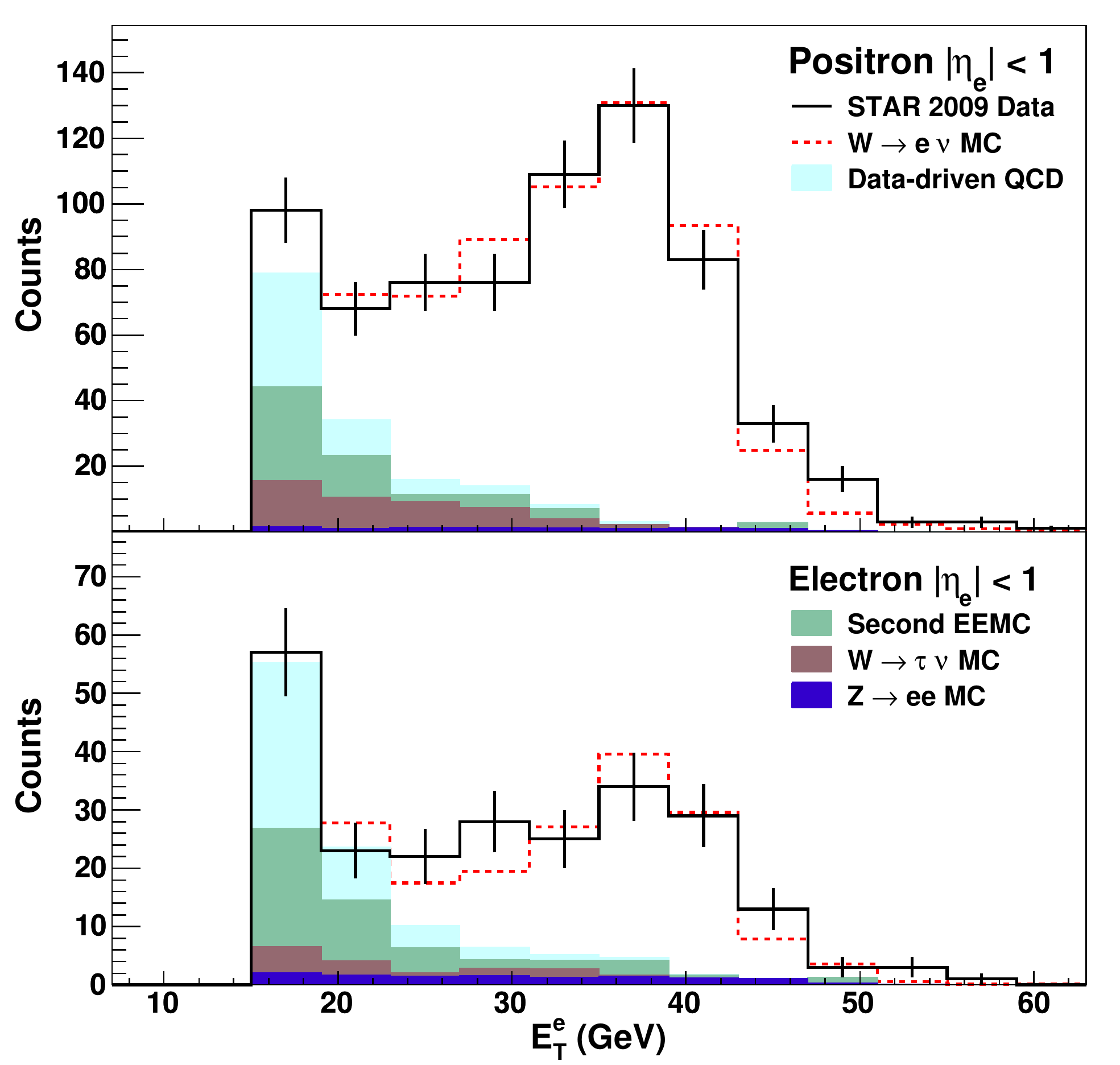}
  \caption{(Color online) $\EeT$ distribution of \Wpl\ (top) and \Wmi\ (bottom) candidate events, background components, and \Wtoenu\ MC signal for comparison.  Note the factor of two difference in the vertical scales.}
  \label{Fig:wJacob}
\end{figure}

The electroweak background processes considered in this analysis are \Wtaunu\ and \Zee.  Their contributions to the \Wtoenu\ signal yield were estimated using the MC samples described in Sec.~\ref{sec:data}.  
\Wtaunu\ events, where the $\tau$ decays leptonically (\textit{i.e.} $\tau \to e\nu\bar{\nu})$, contain an isolated \epm\ with a large missing energy opposite in azimuth, similar to the \Wtoenu\ signal.  However, the \epm\ which comes from the $\tau$ decay must share the energy of the $\tau$ with the two secondary neutrinos, and thus it has a much lower $\EeT$ on average than those \epm\ which come directly from a \Wvar\ decay.  Therefore, the \Wtaunu\ background contributions are largest at low $\EeT$, as can be seen in Fig.~\ref{Fig:wJacob}.
\Zee\ events can contaminate the \Wvar\ signal when one of the decay leptons escapes detection, either from a detector inefficiency or by emission into an uninstrumented region of phase space.  Unlike the other background sources described in this section, the \Zee\ background yield is approximately constant in $\EeT$, resulting in a significant contribution to the total background, even though the cross section is small compared to other processes.  Table \ref{Table:bkgd} lists each of the background processes and its estimated contribution to the \Wvar\ yield for candidates with $\EeT \GREATER 25~\GeV$.  The uncertainties for these electroweak background components are due to the statistical uncertainty of the MC calculation and the uncertainty in the normalization of the MC samples to the integrated luminosity of the data. 

The \STAR\ detector has only one \EEMC, resulting in missing calorimetry acceptance for the pseudorapidity region $-2\LESS\eta\LESS-1.09$ compared to the positive pseudorapidity portion of the detector.  If the isolation cone of $\DELTA\RM{R}\LESS0.7$ around an \epm\ candidate overlaps with this missing acceptance, or a jet opposite in azimuth of an \epm\ candidate falls within this acceptance, background QCD events may satisfy all the \Wtoenu\ selection requirements.  This contamination of the \Wvar\ yield, referred to as the `second \EEMC' background, was determined by repeating the \Wvar\ signal selection a second time, with the \EEMC\ towers excluded from the isolation ratio, $\EeT/\ET^{\DELTA\RM{R}\LESS0.7}$, and the reconstruction of jets summed in the $\vec{p}_{T}^{~bal}$ vector.  The events which satisfy the requirements of this second pass analysis (without the \EEMC), but fail the nominal requirements described in Secs.~\ref{subsec:epm_isolation} and \ref{subsec:Wsignal} are a direct measure of the background rejected by the \EEMC.  Moreover, these events also estimate the amount of background that would have been rejected by a second \EEMC.  

While this sample of second \EEMC\ background is expected to be predominantly the result of QCD processes, it does contain a small amount of \Zee\ contamination as well.  Because background from the \Zee\ process was already taken into account separately, the \Zee\ MC sample was used to remove any contamination from \Zee\ processes in the second \EEMC\ background distribution, to avoid double-counting.  The uncertainty on the second \EEMC\ background is the statistical uncertainty of the events vetoed by the \EEMC\ and the systematic uncertainty in the normalization of \Zee\ contamination which was subtracted using the \Zee\ MC.  

The remaining contribution to the background is predominantly from QCD $2 \to 2$ processes in which one jet fragments such that it satisfies our \epm\ candidate requirements, while all other jets escape detection outside the $|\eta|\LESS2$ acceptance.  This component of the background was estimated using a data-driven QCD background distribution as a function of $\EeT$, which is obtained by selecting events which satisfy all the isolated \epm\ candidate criteria, but have a signed $P_{T}$-balance $\LESS 15~\GeVc$.  Similar to the way the second \EEMC\ background was corrected, contributions to the data-driven background distribution from the \Zee\ process were removed using the \Zee\ MC sample, to avoid double-counting the \Zee\ background.

The data-driven QCD background distribution was then normalized to the remaining \Wtoenu\ candidate signal distribution after the \Wtaunu, \Zee, and second \EEMC\ background components had been removed.  The normalization was determined over the range 15~\LESS~\EeT~\LESS~19~\GeV, and accounts for the possibility of true \Wvar\ signal events in this region using the \Wtoenu\ MC.  The systematic uncertainty of this data-driven QCD background contribution was estimated by varying the data-driven background distribution and the \EeT\ region over which the distribution was normalized.  Twenty different background distributions were obtained by varying the cut on the signed $P_T$-balance variable from 5 to 25 GeV/c in steps of 1 GeV/c. The twenty background distributions were then fit to the signal, as described above, using three different normalization regions (15~$\LESS~\EeT~\LESS$~17,~19,~and~21~\GeV), resulting in sixty different normalized background distributions.   The systematic uncertainty in each \EeT\ bin was taken to be the largest deviation among these sixty distributions from the nominal value.

\begin{table}[!ht]
  \renewcommand{\arraystretch}{1.3}
  \begin{tabular}{|c|c|c|}
    \hline 
    & \Wptoenu\ & \Wmtoenu\ \\
    \hline
    \Wtaunu\           &  13.4 $\pm$ 1.7 $\pm$ 3.2 & 3.3 $\pm$ 0.8 $\pm$ 0.8 \\
    \Zee\              &  7.3  $\pm$ 0.4 $\pm$ 1.7 & 7.3 $\pm$ 0.4 $\pm$ 1.7 \\
    Second \EEMC\      &  9.1  $\pm$ 3.0 $\pm$ 0.5 & 9.2 $\pm$ 3.0 $\pm$ 0.4 \\
    Data-driven QCD    &  7.0  $\pm$ 0.6 $^{+2.3}_{-1.6}$ & 5.8 $\pm$ 0.5 $^{+2.6}_{-1.2}$ \\
    \hline
    Total              &  36.6 $\pm$ 3.5 $^{+5.4}_{-5.2}$ & 25.8 $\pm$ 3.2 $^{+3.6}_{-2.8}$ \\
    \hline
  \end{tabular}
  \caption{Summary of background event contributions to the \Wtoenu\ yield and their uncertainties for candidates with $\EeT\GREATER25~\GeV$ and $|\eta_e|\LESS1$.}
  \label{Table:bkgd}
\end{table}

\begin{figure}[!ht]
  \includegraphics[width=1.0\columnwidth]{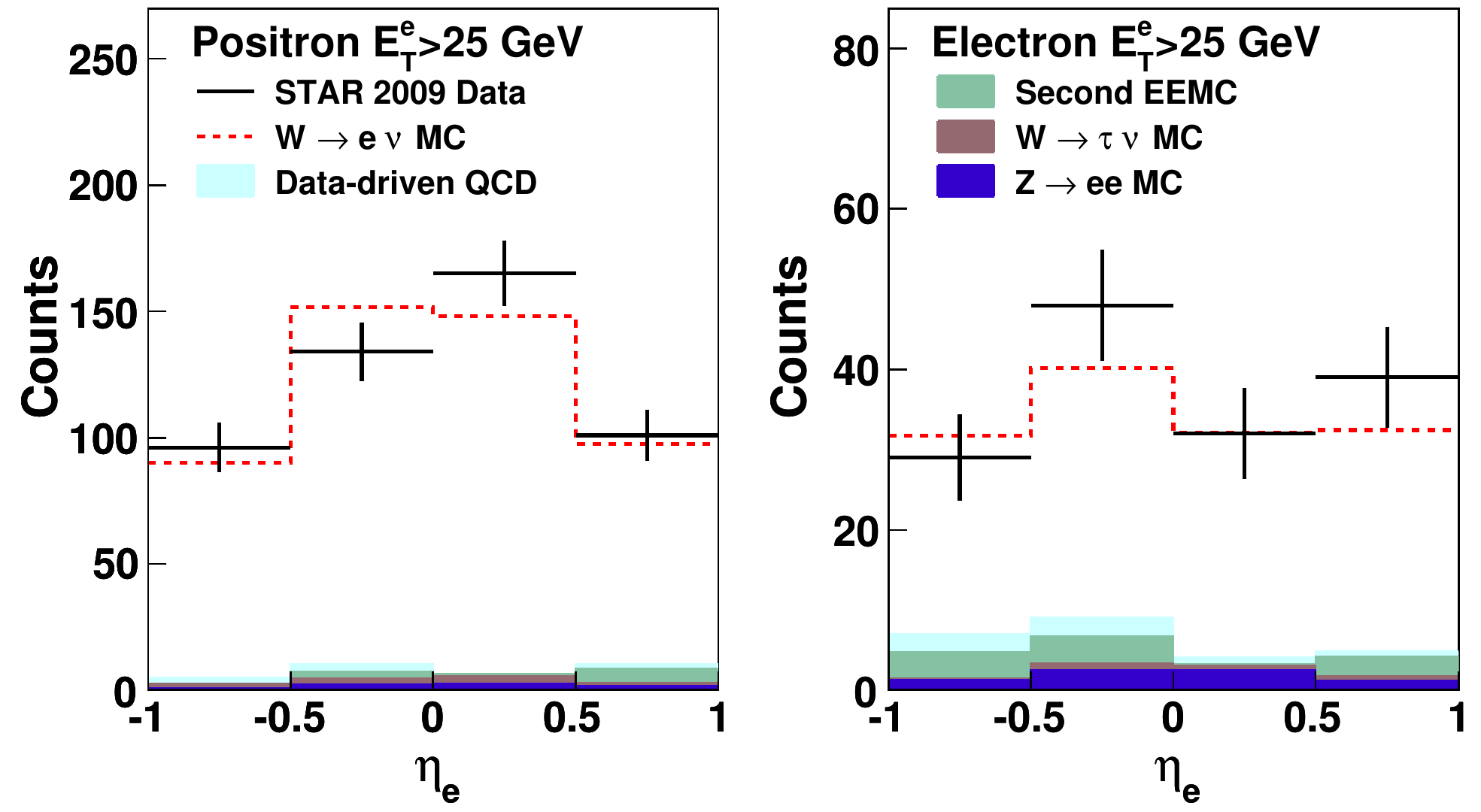}
  \caption{(Color online) Lepton pseudorapidity distribution of \Wpl\ (left) and \Wmi\ (right) candidate events, background components, and \Wtoenu\ MC signal for comparison.}
  \label{Fig:wEta}
\end{figure}

The charge-separated $\EeT$ distributions of \Wpmtoenu\ candidates satisfying all the selection criteria described in Secs.~\ref{subsec:epm_isolation} and \ref{subsec:Wsignal} are shown in Fig.~\ref{Fig:wJacob}.  Also shown here are the contributions from the different backgrounds discussed in this section and the \Wtoenu\ signal MC distribution.  A $\chi^2$ test of homogeneity comparing the data and the sum of background components and \Wtoenu\ signal MC (dashed line) $\EeT$ spectra results in a $\chi^2$ value of 9.5 and 6.9 for the $\Wpl$ and $\Wmi$, respectively.  For 12 degrees of freedom this results in a 66\% and 86\% probability, respectively, to obtain a larger $\chi^2$.  This indicates a good agreement between data and MC and further validates the procedure used in the background estimation described in this section.  The \epm\ pseudorapidity distributions are shown in Fig.~\ref{Fig:wEta}, where the background contributions were found independently for each $\eta_e$ bin using the methods described above.  Again, good agreement is found between the data and the sum of the \Wtoenu\ signal MC and background components. 

\subsection{\label{subsec:Zback}\texorpdfstring{$\bm{\Zvar}$ Background Estimation}{Z Background Estimation}}

The background for the \Zee\ signal is expected to be very small due to the coincidence requirement of a pair of oppositely charged, high $\ET$, isolated $e^+$ and $e^-$.  Background contributions from electroweak processes were estimated using the MC samples described in Sec.~\ref{sec:data}.  Within the defined mass window to be used for the cross section ($70\LESS \mee\LESS110~\GeVcc$), the background contributions were determined to be 0.1 $^{+0.3}_{-0.1}$ events from \Wtoenu\ and negligible from the other \Zvar\ decay channels.  The \Wtoenu\ background uncertainty was estimated using the 68\% C.L. interval of the unified statistical approach described in Ref.~\cite{Feldman:1997qc}.

An accurate data-driven estimate of the QCD background is difficult to obtain for the \Zvar\ signal due to the limited statistics of the data set.  One method for estimating the background is to determine the number of \epm\ pairs that satisfy all the \Zee\ signal criteria other than the opposite charge-sign requirement.  However, no same charge-sign pairs were observed in the data, therefore, the QCD background was found to be consistent with zero.  An upper bound on the QCD background systematic uncertainty was estimated to be 1.3 events using a 68\% C.L. interval \cite{Feldman:1997qc}.

\section{\label{sec:xsec}\texorpdfstring{The $\bm{\Wvar}$ and $\bm{\Zvar}$ Cross Sections}{The W and Z Cross Sections}}

The \Wvar\ and \Zvar\ production cross sections were measured from the sample of events which satisfy the fiducial and kinematic requirements of this analysis.  As stated previously, only \epm\ candidates at mid-rapidity ($|\eta_e|\LESS1$) were considered in this analysis.  Candidates for the \Wvar\ analysis must have $\EeT\GREATER25~\GeV$, and for the \Zvar\ analysis we required that both $e^+$ and $e^-$ have $\EeT\GREATER15~\GeV$ and $70\LESS \mee\LESS110~\GeVcc$.  The cross sections measured within these constraints are defined as the fiducial cross sections, and can be written as:
\begin{equation}
  \sigma_{\Wvar(\Zvar)}^{fid} \cdot \BR(\Wvar(\Zvar)\to e\nu(ee)) = \frac{N^{obs}_{\Wvar(\Zvar)} - N^{bkgd}_{\Wvar(\Zvar)}}{L \cdot \epsilon^{tot}_{\Wvar(\Zvar)}}
  \label{eqn:xSecFidW}
\end{equation}
where
\begin{itemize}
\item $N^{obs}_{\Wvar(\Zvar)}$ is the number of observed $\Wvar(\Zvar)$ candidates within the defined kinematic acceptance, which satisfy all the selection criteria described in Sec.~\ref{sec:signal},
\item $N^{bkgd}_{\Wvar(\Zvar)}$ is the total number of $\Wvar(\Zvar)$ background events within the defined kinematic acceptance described in Sec.~\ref{sec:background},
\item $\epsilon^{tot}_{\Wvar(\Zvar)}$ is the total efficiency correction described in Sec.~\ref{subsec:effic} below,
\item and $L$ is the integrated luminosity of the data set discussed in Sec.~\ref{sec:data}.
\end{itemize}

To determine the total production cross sections, it is necessary to apply acceptance correction factors, $A_{\Wvar(\Zvar)}$, to the fiducial cross sections defined above, to account for the fiducial and kinematic constraints imposed in the analysis.  The total production cross sections are then defined via the relations
\begin{equation}
  \sigma_{\Wvar}^{tot} \cdot \BR(\Wtoenu) = \frac{\sigma_{\Wvar}^{fid} \cdot \BR(\Wtoenu)}{A_{\Wvar}}
\label{eqn:xSecTotW}
\end{equation}
\begin{equation}
\sigma_{\Zvar}^{tot} \cdot \BR(\Zee) = \frac{\sigma_{\Zvar}^{fid} \cdot \BR(\Zee)}{A_{\Zvar}}.
\label{eqn:xSecTotZ}
\end{equation}
The determination of the acceptance corrections necessary to extract the total production cross sections is discussed in Sec.~\ref{subsec:accept}.

\subsection{\label{subsec:effic}The Efficiency Correction Factors}

The efficiency corrections were obtained using the \Wtoenu\ and \Zee\ \PYTHIA\ MC samples described in Sec.~\ref{sec:data}.  Only the subset of events from the MC samples which satisfy the acceptance conditions for the fiducial cross sections were used in the efficiency calculations, as the acceptance correction is accounted for separately in the definition of the total cross section.  

The total efficiency can be factorized into four conditional efficiency terms, written as:
\begin{equation}
  \epsilon_{\Wvar(\Zvar)}^{tot} = \epsilon_{\Wvar(\Zvar)}^{trig} \cdot \epsilon_{\Wvar(\Zvar)}^{vert} \cdot \epsilon_{\Wvar(\Zvar)}^{trk} \cdot \epsilon_{\Wvar(\Zvar)}^{algo} .
  \label{eqn:effic}
\end{equation}
The values for each of the terms in Eq.~\ref{eqn:effic} are listed in Table \ref{Table:effic}, along with their uncertainties, for the \Wpl, \Wmi, and \Zvar\ signals.  The remainder of this section describes how those values were obtained.
  
The trigger efficiency, $\epsilon^{trig}$, is the fraction of MC signal events which satisfy the online trigger condition defined in Sec.~\ref{sec:data}.  This was determined by emulating the trigger condition used online in the MC.  Due to the relatively wide \zvertex\ distribution of our data sample, some candidates may satisfy the $|\eta_e|\LESS1$ kinematic condition at the MC generator level, but will fall outside the acceptance of the \BEMC.  This was observed in the \Wvar\ analysis as an $\EeT$-dependent trigger efficiency due to the correlation of the $\EeT$ and $\eta_e$ of the decay \epm.  An $\EeT$-dependent trigger efficiency correction was therefore used in the computation of the \Wpm\ cross sections.  This effect also leads to a notably smaller average \Wmi\ trigger efficiency relative to \Wpl, as the $\eta_e$ distribution is expected to be peaked more strongly at zero for the \Wpl\ candidates than \Wmi, which is consistent with Fig.~\ref{Fig:wEta}.  To estimate the uncertainty on $\epsilon^{trig}$, the \BEMC\ energy scale was varied by its uncertainty of $\pm$3.6\%.  Because the offline kinematic requirement of $\EeT\GREATER25~\GeV$ was significantly larger than the trigger threshold of 13$~\GeV$, for this analysis we observed only small variations in the trigger efficiency due to the uncertainty of the \BEMC\ energy calibration.  

The vertex efficiency, $\epsilon^{vert}$, is defined as the fraction of events satisfying the trigger which contain a reconstructed primary vertex within the fiducial cut of $|\zvertex|\LESS100~\cm$, as described in Sec.~\ref{sec:signal}.  The tracking efficiencies for the \Wvar\ and \Zvar\ decay $\epm$s are defined as follows.  For \Wvar\ events with a reconstructed primary vertex, $\epsilon_{W}^{trk}$ is the efficiency for reconstructing a single \TPC\ track which satisfies the track requirements in Sec.~\ref{subsec:epm_isolation}, however for \Zee\ events the tracking efficiency, $\epsilon_{Z}^{trk}$, is the efficiency for reconstructing \textit{two} \TPC\ tracks satisfying those conditions.  In comparing the reconstructed \TPC\ track 1/\pT\ distributions between data and MC, a slightly worse resolution was seen in the data.  This was accounted for by re-weighting the MC distributions to match the data.  The uncertainty on the tracking efficiency was estimated from the error in this re-weighting resulting from the limited statistics of the data distribution.

Finally, the algorithm efficiency, $\epsilon^{algo}$, is the fraction of events with one (two) reconstructed \epm\ candidate \TPC\ tracks, which satisfy the remaining \Wvar\ (\Zvar) selection criteria.  As discussed in Sec.~\ref{sec:signal}, these remaining selection criteria include reconstruction of \BEMC\ clusters, matching extrapolated track and cluster positions, isolation requirements, and finally the signed $P_T$-balance and pair of opposite charge-sign candidate requirements for \Wvar\ and \Zvar\ events, respectively.  A weak $\EeT$ dependence was observed in the algorithm efficiency for the \Wtoenu\ MC due mainly to the efficiency of the $\EeT/\ET^{\DELTA\RM{R}\LESS0.7}$ isolation cut being reduced at low $\EeT$.  Thus, an $\EeT$-dependent algorithm efficiency correction was used in the computation of the \Wpm\ cross sections.  The uncertainty on $\epsilon^{algo}$ was determined by varying the \BEMC\ scale uncertainty, as was done for the trigger efficiency. 

\begin{table}[!ht]
  \renewcommand{\arraystretch}{1.3}
  \begin{tabular}{|c|c|c|c|}
    \hline 
    & \Wptoenu\ & \Wmtoenu\ & \Zee\ \\
    \hline 
    $\epsilon^{trig}$    & 0.857 $\pm$ 0.007 & 0.825 $\pm$ 0.007 & 0.968 $\pm$ 0.006 \\
    $\epsilon^{vert}$  & 0.881 $\pm$ 0.005 & 0.886 $\pm$ 0.006 & 0.938 $\pm$ 0.006 \\
    $\epsilon^{trk}$   & 0.741 $\pm$ 0.030 & 0.748 $\pm$ 0.031 & 0.511 $\pm$ 0.032 \\
    $\epsilon^{algo}$    & 0.892 $\pm$ 0.024 & 0.892 $\pm$ 0.024 & 0.730 $\pm$ 0.024 \\
    \hline
    $\epsilon^{tot}$     & 0.498 $\pm$ 0.026 & 0.488 $\pm$ 0.026 & 0.338 $\pm$ 0.024 \\
    \hline 
  \end{tabular}
  \caption{Summary of conditional efficiency correction factors included in Eq.~\ref{eqn:effic}.  The average values for the trigger and algorithm efficiencies for the \Wpm\ analysis are given here, however an $\EeT$-dependent correction was used for the measured cross section, as described in the text.}
  \label{Table:effic}
\end{table}

\subsection{\label{subsec:xSecFid}The Measured Fiducial Cross Sections}

The fiducial cross sections are calculated according to Eq.~\ref{eqn:xSecFidW}, and the measured values are summarized in Tables \ref{Table:xSecFidW} and \ref{Table:xSecFidZ} for \Wpm\ and \Zvar\, respectively.  The dominant uncertainty for both the \Wpl\ and \Wmi\ cross sections comes from the systematic uncertainty in the measured luminosity of the data sample.  The \Zvar\ cross section measurement, however, is currently dominated by the statistical uncertainty. 

\begin{table}[!ht]
  \renewcommand{\arraystretch}{1.3}
  \begin{tabular}{|c|c|c|c|c|c|c|c|c|}
    \hline 
    & \multicolumn{4}{c|}{$\Wptoenu$} & \multicolumn{4}{c|}{$\Wmtoenu$} \\
    \hline
    & value & stat & syst & lumi & value & stat & syst & lumi \\
    \hline 
    $N^{obs}$         & 496 & 22.3 & - & - & 148 & 12.2 & - & - \\
    $N^{bkgd}$        & 36.6 & 3.5 & $^{+5.4}_{-5.2}$ & - & 25.8 & 3.2 & $^{+3.6}_{-2.8}$ & - \\
    $\epsilon^{tot}$  & 0.498 & 0.006 & 0.025 & - & 0.488 & 0.007 & 0.025 & - \\
    $L~(\pbinv)$               & 13.2 & 0.2 & - & 1.7 & 13.2 & 0.2 & - & 1.7 \\
    \hline
    $\sigma^{fid}~(\pb)$    & 70.0 & 3.5 & 3.5 & 9.1 & 19.2 & 2.1 & 1.1 & 2.5 \\
    \hline 
  \end{tabular}
  \caption{Summary of input and measured values for the $\Wtoenu$ fiducial cross sections, with their statistical, systematic, and luminosity uncertainties.  As noted in the text, an $\EeT$-dependent efficiency correction factor is used for the cross section measurement, and only the average value is shown here.}
  \label{Table:xSecFidW}
\end{table}

\begin{table}[!ht]
  \renewcommand{\arraystretch}{1.3}
  \begin{tabular}{|c|c|c|c|c|}
    \hline 
    & \multicolumn{4}{c|}{$\Zee$} \\
    \hline
    & value & stat & syst & lumi \\
    \hline 
    $N^{obs}$         & 13 & 3.6 & - & - \\
    $N^{bkgd}$        & 0.1 & 0.1 & $^{+1.3}_{-0.0}$ & - \\
    $\epsilon^{tot}$  & 0.338 & 0.012 & 0.021 & - \\
    $L~(\pbinv)$     & 13.2 & 0.2 & - & 1.7 \\
    \hline
    $\sigma^{fid}~(\pb)$    & 2.9 & 0.8 & $^{+0.2}_{-0.3}$ & 0.4 \\
    \hline 
  \end{tabular}
  \caption{Summary of input and measured values for the \Zee\ fiducial cross section, with their statistical, systematic, and luminosity uncertainties.}
  \label{Table:xSecFidZ}
\end{table}

\subsection{\label{subsec:accept}The Acceptance Correction Factors}

As stated previously, to determine the total cross sections, acceptance correction factors $A_{\Wvar(\Zvar)}$, must be used to account for the fiducial and kinematic acceptance requirements of the analysis, which are defined at the beginning of Sec.~\ref{sec:xsec}.  $A_{\Wvar(\Zvar)}$ were calculated using the \FEWZ\ program \cite{Melnikov:2006kv}, which provides cross section calculations for \Wvar\ and \Zvar\ boson production up to NNLO in pQCD.  Table~\ref{Table:accept} lists the values of the acceptance factors using the MSTW 2008 \cite{Martin:2009iq} and CTEQ 6.6 \cite{Nadolsky:2008zw} parton distribution function sets.  The nominal values for the acceptance corrections, used in the total cross section measurements, were taken from the next-to-leading order (\NLO) calculation using the MSTW08 PDF set.  Theoretical uncertainties in the calculation of these factors arise from several sources, including differences between PDF sets, uncertainties within a PDF set, and uncertainties in the modeling of the production process.

\begin{table}[!ht]
  \centering
  \renewcommand{\arraystretch}{1.3}
  \begin{tabular}{|c|c|c|c|}
    \hline
    & $A_{\Wpl}$ & $A_{\Wmi}$ & $A_{\Zvar}$ \\
    \hline
    LO   MSTW08    & 0.591 & 0.444 & 0.377 \\
    \NLO\  MSTW08  & 0.597 & 0.444 & 0.378 \\
    \NNLO\ MSTW08  & 0.603 & 0.435 & 0.385 \\
    \hline
    \NLO\  CTEQ 6.6   & 0.592 & 0.430 & 0.370 \\
    \hline
  \end{tabular}
  \caption{ Summary of acceptance values calculated with the $\FEWZ$ program.  The \NLO\ MSTW08 values are used for the total cross section calculations in Sec.~\ref{subsec:xSecTot}. }
  \label{Table:accept}
\end{table}

The uncertainty due to differences between PDF sets was taken to be the difference between the CTEQ 6.6 and MSTW08 acceptance values at NLO.  Both groups provide error eigenvector PDF sets which were used to estimate the acceptance uncertainty, at the 90\% confidence level, within each set.  The average of the CTEQ 6.6 and MSTW08 error eigenvector uncertainty was taken to be the uncertainty due to the PDF itself.  Finally, the uncertainty in the modeling of the production process was estimated by comparing the acceptance values from calculations with different orders of QCD corrections, using the MSTW08 PDF set.  The maximum difference from the nominal value (\NLO\ MSTW08) was taken as this final uncertainty contribution.  Table \ref{Table:acceptUncert} summarizes the contributions to the uncertainties in the acceptance values.  The individual contributions were added in quadrature to determine the total uncertainty for each acceptance factor.  The $A_{\Wmi}$ uncertainties are significantly larger than those for $A_{\Wpl}$, driven primarily by the PDF-related errors.  This is expected, due to the larger uncertainties in the $\bar{u}$ and $d$ quark PDFs with respect to those of the $\bar{d}$ and $u$ quarks.

\begin{table}[!ht]
  \centering
  \renewcommand{\arraystretch}{1.3}
  \begin{tabular}{|c|c|c|c|}
    \hline
    & $\delta A_{\Wpl} (\%)$ & $\delta A_{\Wmi} (\%)$ & $\delta A_{\Zvar} (\%)$ \\
    \hline
    Difference between PDFs    & 1.0 & 3.2 & 2.1 \\
    MSTW08 \NLO\ error PDFs    & 0.9 & 2.7 & 1.2 \\
    CTEQ 6.6 \NLO\ error PDFs  & 0.9 & 4.5 & 1.8 \\
    Calculation Order          & 1.0 & 2.0 & 1.9 \\
    \hline
    Total                      & 1.7 & 5.2 & 3.2 \\
    \hline
  \end{tabular}
  \caption{ Summary of the relative uncertainties in the acceptance correction factors, $A_{\Wvar(\Zvar)}$, as computed by the \FEWZ\ program. }
  \label{Table:acceptUncert}
\end{table}

\subsection{\label{subsec:xSecTot}The Measured Total Cross Sections}

The total cross sections are calculated according to Eqs. \ref{eqn:xSecTotW} and \ref{eqn:xSecTotZ}, by dividing the measured fiducial cross sections by the acceptance correction factors determined in the previous section.  The results for $\pp \to \Wpm$ total production cross sections at \sqrts\ = 500~\GeV\ are the following:
\begin{center}
  $\sigma_{\Wpl}^{tot} \cdot \BR(\Wptoenu)$ = 117.3 $\pm$ 5.9(stat) \\ $\pm$ 6.2(syst) $\pm$ 15.2(lumi) pb,

  \bigskip

  $\sigma_{\Wmi}^{tot} \cdot \BR(\Wmtoenu)$ = 43.3 $\pm$ 4.6(stat) \\ $\pm$ 3.4(syst) $\pm$ 5.6(lumi) pb. 
\end{center}
The result for the $\pp \to \Zgam$ total production cross section at \sqrts\ = 500~\GeV\ in the invariant mass range of $70\LESS \mee\LESS110~\GeVcc$ is 
\begin{center}
  $\sigma_{\Zgam}^{tot} \cdot \BR(\Zgtoee)$ = 7.7 $\pm$ 2.1(stat) \\ $^{+0.5}_{-0.9}$(syst) $\pm$ 1.0(lumi) pb.
\end{center} 

\begin{figure}[!ht]
  \includegraphics[width=1.0\columnwidth]{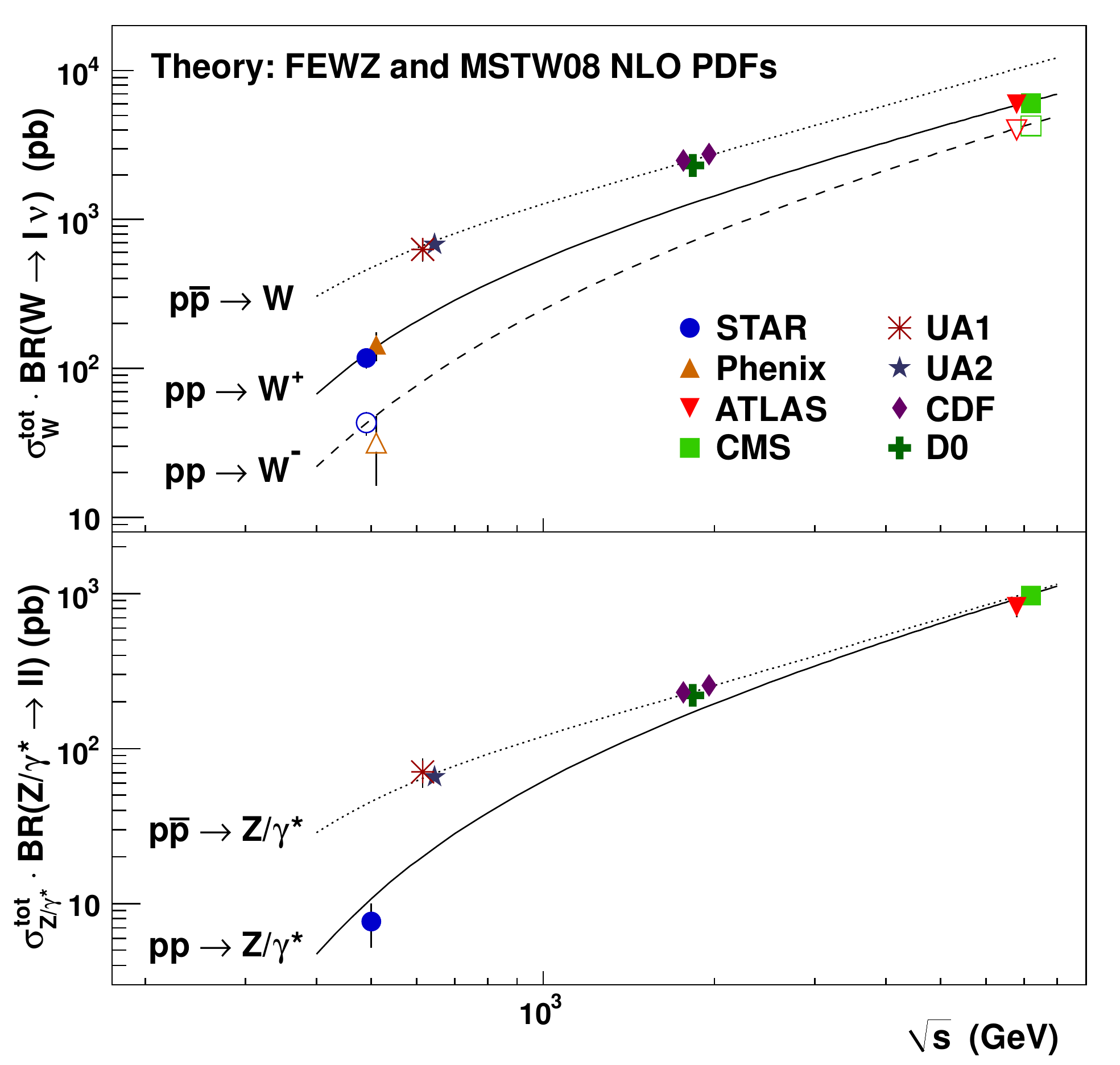}
  \caption{(Color online) Measurements of \Wvar\ and \Zvar\ total cross sections times branching ratio versus center-of-mass energy.  For the \Wvar\ cross sections in \pp\ collisions, the closed symbols represent \Wpl\ and the open symbols represent \Wmi.  The theory curves are from the \FEWZ\ program at \NLO\ using the MSTW08 PDF set.}
  \label{Fig:xSecBR}
\end{figure}

Figure \ref{Fig:xSecBR} shows the measured total cross sections, multiplied by the respective branching ratios, in comparison with the theoretical predictions at \NLO\ from the \FEWZ\ program using the MSTW08 PDF set.  Measurements from other experiments at the $\rm{Sp \bar{p} S}$, Tevatron, RHIC, and LHC are also shown as a function of \sqrts\ for comparison. 

Theoretical predictions for the production cross sections computed by the \FEWZ\ \cite{Melnikov:2006kv} and fully resummed \RHICBOS\ \cite{Nadolsky:2003ga} calculations are shown in Table \ref{Table:xSecTheory}.  The theoretical uncertainties were determined for the \FEWZ\ predictions using the 90\% confidence level error eigenvector PDF sets; error eigenvector sets are not provided for the \RHICBOS\ calculation.  Variations in the strong coupling constant, $\alpha_s$, from the associated error PDF sets were considered as well, but the uncertainties were found to be negligible compared to the uncertainties from the PDFs.  The theoretical predictions agree well with the measured cross sections within the theoretical and experimental uncertainties.  Interestingly, differences between the MSTW08 and CTEQ 6.6 PDF sets result in significant differences in the predicted cross sections at \NLO.  

\begin{table}[!ht]
  \centering
  \renewcommand{\arraystretch}{1.3}
  \begin{tabular}{|c|c|c|c|}
    \hline
    & $\sigma^{tot}_{\Wpl} (\pb)$ & $\sigma^{tot}_{\Wmi} (\pb)$ & $\sigma^{tot}_{\Zvar}\cdot (\pb)$ \\
    \hline
    \NLO\  MSTW08    & 132.4 $\pm$ 9.0 & 45.7 $\pm$ 3.6 & 10.8 $\pm$ 0.8 \\
    \NNLO\ MSTW08    & 136.7 $\pm$ 9.5 & 48.1 $\pm$ 3.0 & 11.2 $\pm$ 0.8 \\
    \hline
    \NLO\  CTEQ 6.6  & 121.8 $\pm$ 8.8 & 41.1 $\pm$ 4.3 & 9.8  $\pm$ 0.8 \\
    \hline
    Ressum.  CTEQ 6.6  & 121.1 & 39.9 & - \\
    \hline
  \end{tabular}
  \caption{ Summary of total cross section (times branching ratio) theoretical predictions at \sqrts\ = 500~\GeV\ calculated with the \FEWZ\ and \RHICBOS\ programs.  The \Zgam\ values are defined within the invariant mass range of $70\LESS \mee\LESS110~\GeVcc$. }
  \label{Table:xSecTheory}
\end{table}

\section{\label{sec:ratio}\texorpdfstring{The $\bm{\Wvar}$ Cross Section Ratio}{The W Cross Section Ratio}}

The $\Wvar$ cross section ratio is defined as
\begin{equation}
  R_\Wvar=\frac{\sigma^{fid}_{\Wpl}}{\sigma^{fid}_{\Wmi}} = \frac{N^{obs}_{\Wpl} - N^{bkgd}_{\Wpl}}{N^{obs}_{\Wmi} - N^{bkgd}_{\Wmi}} \cdot \frac{\epsilon^{tot}_{\Wmi}}{\epsilon^{tot}_{\Wpl}}.
\end{equation}
If the small contributions from strange quarks are neglected, this ratio should be equal to~\cite{Peng:1995ba}
\begin{equation}
  R_\Wvar=\frac{u(x_1)\bar{d}(x_2)+\bar{d}(x_1)u(x_2)}{\bar{u}(x_1)d(x_2)+d(x_1)\bar{u}(x_2)}. 
  \label{eqn:R_W}
\end{equation}
Measurements of the cross section ratio should therefore be sensitive to the flavor asymmetry of the antiquark sea in the Bjorken-$x$ range $0.1\LESSAPPROX\IT{x}\LESSAPPROX0.3$ probed at RHIC.  Drell-Yan experiments \cite{Baldit:1994jk,Towell:2001nh} have measured a large asymmetry in this $x$ range, and precision measurements of $R_W$ at RHIC can provide independent constraints on the flavor asymmetry which are free from the assumption of charge symmetry required in Drell-Yan.  Measurements of the lepton charge asymmetry at the LHC \cite{Aad201131,Chatrchyan:2011jz} provide similar constraints on the quark and antiquark PDFs, though at significantly lower $x$ due to the much higher energy of the collisions.

The \Wvar\ cross section ratio was measured in two $|\eta_e|$ regions, as this coarsely constrains the $x$ of the partons involved in the \Wvar\ production.  In each $|\eta_e|$ bin, the fiducial cross sections were computed using the same procedures described in Sec.~\ref{sec:xsec}, where the background and efficiencies were separately calculated for each charge and $|\eta_e|$ bin.  The luminosity, and its sizable uncertainty, cancel in the cross section ratio, significantly reducing the systematic uncertainty, with respect to the \Wpl\ and \Wmi\ cross sections independently. 

\begin{table}[!ht]
  \centering
  \renewcommand{\arraystretch}{1.3}
  \begin{tabular}{|c|c|}
    \hline
    & $R_\Wvar \pm$ (stat) $\pm$ (syst) \\
    \hline
    $|\eta_e|\LESS0.5$           & 4.3 $\pm$ 0.7 $\pm$ 0.3 \\
    $0.5\LESS|\eta_e|\LESS1.0$   & 2.9 $\pm$ 0.5 $\pm$ 0.2 \\
    \hline
  \end{tabular}
  \caption{(Color online) Measurements of the \Wvar\ cross section ratio, $R_\Wvar$, for the two \epm\ pseudorapidity bins. }
  \label{Table:xSecRatio}
\end{table}

\begin{figure}[!ht]
  \includegraphics[width=1.0\columnwidth]{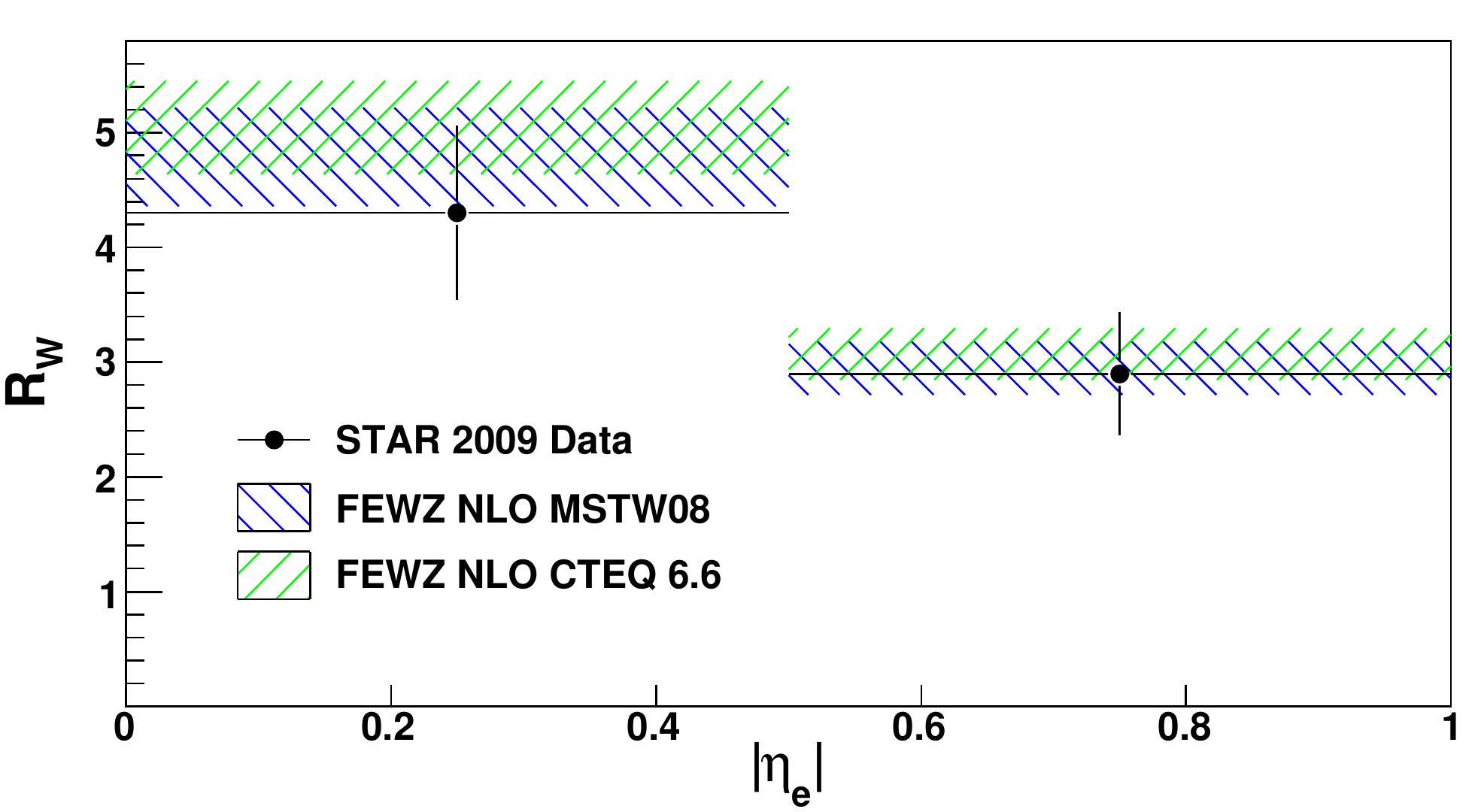}
  \caption{(Color online) \Wvar\ cross section ratio, $R_\Wvar$, for the two \epm\ pseudorapidity bins.  Theory calculations at NLO from the \FEWZ\ program using the MSTW08 and CTEQ 6.6 PDF sets (with 90\% confidence level error eigenvector uncertainties) are shown for comparison.}
  \label{Fig:xSecRatio}
\end{figure}

Our results for the measured cross section ratio are listed in Table \ref{Table:xSecRatio}.  Figure \ref{Fig:xSecRatio} shows the cross section ratio as a function of $|\eta_e|$, where the statistical and systematic uncertainties of the data have been added in quadrature.  Also displayed in Fig.~\ref{Fig:xSecRatio} are theoretical calculations of the cross section ratio computed with the \FEWZ\ program at \NLO.  Both the MSTW08 and CTEQ 6.6 PDF sets were used to compute the ratio; the error bands shown are the 90\% confidence level error eigenvector uncertainties.  The predictions agree with the measured values within the large uncertainties, which are dominated by the statistical precision of the \Wmi\ yield.

\section{\label{sec:summary}Summary}

We have presented measurements of the \Wptoenu, \Wmtoenu, and \Zgtoee\ production cross sections in proton-proton collisions at $\sqrts = 500~\GeV$ by the \STAR\ detector at \RHIC.  Theoretical predictions based on pQCD calculations are in good agreement with the measured cross sections.  In addition, a first measurement of the \Wvar\ cross section ratio is presented.  Future high statistics measurements of the \Wvar\ cross section ratio at RHIC will provide a new means of studying the flavor asymmetry of the antiquark sea which is complementary to fixed-target Drell-Yan and LHC collider measurements.

\bigskip

\begin{acknowledgments}

We thank the RHIC Operations Group and RCF at BNL, the NERSC Center at LBNL and the Open Science Grid consortium for providing resources and support. We are grateful to F. Petriello for useful discussions.  This work was supported in part by the Offices of NP and HEP within the U.S. DOE Office of Science, the U.S. NSF, the Sloan Foundation, the DFG cluster of excellence `Origin and Structure of the Universe' of Germany, CNRS/IN2P3, FAPESP CNPq of Brazil, Ministry of Ed. and Sci. of the Russian Federation, NNSFC, CAS, MoST, and MoE of China, GA and MSMT of the Czech Republic, FOM and NWO of the Netherlands, DAE, DST, and CSIR of India, Polish Ministry of Sci. and Higher Ed., Korea Research Foundation, Ministry of Sci., Ed. and Sports of the Rep. of Croatia, and RosAtom of Russia.

\end{acknowledgments}

\bibliography{W-PRD-v1.9}

\end{document}